\documentclass[12pt,tightenlines,showpacs,preprintnumbers,superscriptaddress,nofootinbib]{revtex4-1}
\pdfoutput=1

\usepackage[pdftex]{graphicx}
\usepackage[utf8]{inputenc}
\usepackage[T1]{fontenc}
\usepackage{verbatim}
\usepackage{enumerate}
\usepackage{amssymb} 
\usepackage{amsmath}
\usepackage{color}
\usepackage[normalem]{ulem}
\usepackage[colorlinks=true
,urlcolor=blue
,anchorcolor=blue
,citecolor=blue
,filecolor=blue
,linkcolor=blue
,menucolor=blue
,linktocpage=true
,pdfproducer=medialab
,pdfa=true
]{hyperref}

\newcommand{\df}{\dfrac}

\newcommand{\mh}{m_h}
\newcommand{\ma}{m_A}
\newcommand{\meta}{m_\eta}
\newcommand{\xf}{x_\text{FO}}
\newcommand{\xpf}{x'_\text{FO}}
\newcommand{\n}{n'}
\newcommand{\neqq}{{n'}^\text{eq}}
\newcommand{\RR}[5]{\langle #1#2#3\to #4#5\rangle\left(\n_#1\,\n_#2\,\n_#3-\n_#4\,\n_#5\frac{\neqq_#1\,\neqq_#2\,\neqq_#3}{\neqq_#4\,\neqq_#5}\right)}
\newcommand{\NN}[5]{\langle #1#2#3\to #4#5\rangle}

\newcommand{\PRE}[1]{{#1}} 

\begin{document}

\preprint{ULB-TH/15-15}
\preprint{LAPTH-057/15}

\title{ \PRE{\vspace*{1.5in}} 
Production Regimes for Self-Interacting Dark Matter
\PRE{\vspace*{0.3in}} }

\author{Nicol\'as Bernal}
\affiliation{ICTP South American Institute of Fundamental Research\\
Instituto de F\'isica Te\'orica, Universidade Estadual Paulista, S\~ao Paulo, Brazil \PRE{\vspace*{.2in}} }
\author{Xiaoyong Chu}
\affiliation{International Centre for Theoretical Physics, ICTP\\
Strada Costiera 11, 34014 Trieste, Italy\PRE{\vspace*{.2in}}}
\author{Camilo Garcia-Cely}
\author{Thomas Hambye}
\affiliation{Service de Physique Th\'eorique, Universit\'e Libre de Bruxelles\\ 
Boulevard du Triomphe, CP225, 1050 Brussels, Belgium\PRE{\vspace*{.2in}}}
\author{Bryan Zaldivar}
\affiliation{Service de Physique Th\'eorique, Universit\'e Libre de Bruxelles\\ 
Boulevard du Triomphe, CP225, 1050 Brussels, Belgium\PRE{\vspace*{.2in}}}
\affiliation{LAPTh, Universit\'e de Savoie Mont Blanc\\ 
CNRS, B.P.110, F-74941 Annecy-le-Vieux, France\PRE{\vspace*{.2in}}}

\begin{abstract}
In the context of Self-Interacting Dark Matter as a solution for the small-scale structure problems, we consider the possibility that Dark Matter
could have been produced without being 
in thermal equilibrium with the Standard Model bath. 
We discuss one by one the following various dark matter production regimes of this kind: freeze-in, reannihilation and dark freeze-out.
We exemplify how these mechanisms work in the context of the particularly simple Hidden Vector Dark Matter model. In contrast to scenarios where there is thermal equilibrium with the Standard Model bath, we find two regimes which can easily satisfy all the laboratory and cosmological constraints. These are dark freeze-out with 3-to-2 annihilations and freeze-in via a light mediator. In the first regime, different temperatures in the visible and the Dark Matter sectors allow us to avoid the constraints coming from cosmic structure formation 
as well as the use of non-perturbative couplings to reproduce the observed relic density. For the second regime, different couplings are responsible for Dark Matter
relic density and self-interactions, permitting to surpass BBN, $X$-ray,
CMB and direct detection constraints.

\end{abstract}

\maketitle
\tableofcontents

\newpage

\section{Introduction}
\label{sec:intro} 
The search for non-gravitational interactions of Dark Matter (DM) constitutes one of the main concerns of particle physics today~\cite{2012AnP...524..479B,Profumo:2013yn,Gelmini:2015zpa}.
 Although the three main different DM search strategies (direct detection, indirect detection and collider searches) have not provided any conclusive signal so far,
these strategies offer real possibilities of discovering soon the DM particle.
Yet another strategy is to look for effects associated to possible DM non-gravitational self-interactions in the sky. In fact, the long-standing puzzles of the collisionless cold DM paradigm can be addressed by invoking such interactions. Two examples of such puzzles are the `cusp vs. core'~\cite{Moore:1994yx,Flores:1994gz,Oh:2010mc,Walker:2011zu} and the `too-big-to-fail'~\cite{BoylanKolchin:2011de,Garrison-Kimmel:2014vqa} problems.  These can be alleviated if at the scale of galaxies there exists a large self-scattering cross section, $\sigma_{AA}$, over DM particle mass, $\ma$, in the range $0.1\lesssim\sigma_{AA}/\ma\lesssim10$ cm$^2/$g~\cite{Spergel:1999mh,Wandelt:2000ad,Vogelsberger:2012ku,Rocha:2012jg,Peter:2012jh,Zavala:2012us,Vogelsberger:2014pda,Elbert:2014bma,Kaplinghat:2015aga}.
Nevertheless, the non-observation of an offset between the mass distribution of DM and galaxies in the Bullet Cluster constrains such self-interacting cross section, concretely $\sigma_{AA}/\ma<1.25$ cm$^2$/g at $68\%$~CL~\cite{Clowe:2003tk,Markevitch:2003at,Randall:2007ph}, i.e. around $10^{12}$~pb for a DM of 1\,GeV mass. Similarly, recent observations of cluster collisions  lead to the constraint $\sigma_{AA}/\ma<0.47$~cm$^2$/g at $95\%$~CL~\cite{Harvey:2015hha}.%
\footnote{A $1.62$~kpc offset between DM and stars has been recently claimed in the cluster Abell 3827.
If interpreted solely as an effect of DM self-interaction, it implies a non-vanishing $\sigma_{AA}/\ma$ of the order of $10^{-4}$~cm$^2$/g~\cite{Massey:2015dkw}. However, these results have been reinterpreted using a different kinematical analysis, obtaining $\sigma_{AA}/\ma\sim 1.5$~cm$^2$/g in the case of contact interactions~\cite{Kahlhoefer:2015vua}.
Since these recent results are inconclusive at this moment, we do not consider them in this work. }

Building a consistent model with such a large self-interaction cross section that accounts also for the observed amount of the DM relic density is theoretically not straightforward. 
As a matter of fact, such a cross section value is orders of magnitude larger than the typical `weak interaction' DM annihilation cross section, around 1 pb~\cite{Steigman:2012nb},
which is required in the usual thermal freeze-out scenario. 
In fact, the fast interactions responsible for the self-interactions typically induce strong annihilation cross sections, which naturally lead to 
a negligible DM relic abundance~\cite{Feng:2009hw, Tulin:2012wi} (see also the
discussion in Ref.~\cite{Izaguirre:2015yja}).
Therefore, one needs a mechanism where DM self-interactions are enhanced today and/or where the annihilation cross section is  suppressed in the early Universe.

One possibility is to introduce light mediators to boost DM self-interactions~\cite{Feng:2009hw, Buckley:2009in, Loeb:2010gj, Tulin:2013teo}.
This is based on the fact that the exchange of a light mediator leads to enhancement effects, which are more pronounced today than in the early Universe due to the different DM velocities in these eras. Such a possibility is constrained in multiple ways.
For instance, the mediator itself contributes to the matter relic density if it is very long-lived.
A very light mediator can also contribute to the amount of relativistic degrees of freedom (Dark Radiation) in the early Universe.
Also, such contributions could spoil Big Bang Nucleosynthesis (BBN) or be in conflict with 
Cosmic Microwave Background (CMB) constraints.
Complementarily, unstable mediators which interact significantly with the Standard Model (SM) sector may lead to observable signals in DM direct detection experiments~\cite{Kaplinghat:2013yxa, DelNobile:2015uua, Kainulainen:2015sva} and even at colliders~\cite{Kainulainen:2015sva}.
As a result, this possibility is severely constrained, excluded in many models but still viable in others.

Another possibility is to consider a framework where the freeze-out of the DM particles does not proceed via a usual self-annihilation of DM pairs, but via processes where three DM particles annihilate into two of them~\cite{Carlson:1992fn, Hochberg:2014dra}. In this case, large self-interactions do not lead to too large 3-to-2 annihilations because the latter are naturally phase-space suppressed. Nonetheless, in the most simple cases the relic density and self-interaction that addresses the small-scale problems lead to DM couplings on the verge of non-perturbativity~\cite{Hansen:2015yaa}. 
Another problem associated to the 3-to-2 freeze-out mechanism is that such processes relatively increase the temperature of the DM particles if they are not in kinetic equilibrium with the SM bath in the early Universe~\cite{Carlson:1992fn,Machacek:1994vg} {(See Section~\ref{sec:NoLightMediator} for details)}. In this case, DM particles may end up being too hot to be consistent with structure formation observations, even if DM free streaming length is suppressed by strong self-interactions. 
A possible way out of this problem is to consider a mechanism establishing kinetic equilibrium between the DM and the photons~\cite{Hochberg:2014dra,Choi:2015bya} or another relativistic species~\cite{Bernal:2015bla,Bernal:2015lbl}. Nonetheless, in many cases,
the large couplings to the SM that are needed for establishing the equilibrium lead to the decay of the SM scalar into DM particles, making this possibility tightly constrained.

In the literature, the light mediator and 3-to-2 annihilation possibilities  have essentially been  considered in frameworks where, to
account for the DM relic density, the usual thermal freeze-out mechanism takes place. However, as is well known, this is not the only mechanism that can account for the DM abundance. In particular if DM belongs to a sector which is feebly coupled to the SM sector,
other  DM production mechanisms are also allowed, such as freeze-in~\cite{McDonald:2001vt,Hall:2009bx}, reannihilation~\cite{Cheung:2010gj,Cheung:2010gk,Chu:2011be} or freeze-out in the hidden sector~\cite{Feng:2008mu,Ackerman:mha}. All these mechanisms assume that the DM has not been in chemical and/or kinetic equilibrium with the SM bath.  We analyze here how  DM self-interactions can be accommodated in the framework of these others regimes, both with a light mediator or with a 3-to-2 annihilation scheme.
We show how, for any given model, a systematic scan of the production regimes for self-interacting DM can be done.

To this end we consider the various regimes in the framework of a particularly simple and minimal example model, the Hidden Vector DM setup (HVDM)~\cite{Hambye:2008bq,Hambye:2009fg,Arina:2009uq}. In this model, we find two regimes which allow to fulfill all the aforementioned problems in a simple way. One of them consists of  a freeze-out via 3-to-2 annihilation  processes, in a scenario where the temperature of the Hidden Sector (HS) is smaller than the temperature of the visible sector. The other one invokes freeze-in via a light mediator.\footnote{See also Refs.~\cite{Campbell:2015fra,Kang:2015aqa} for very recent works where the freeze-in mechanisms has also been invoked for DM self-interactions and Ref.~\cite{Boddy:2014yra} for a more involved framework where the HS freeze-out mechanism has been used.}

In order to perform all discussions based on a concrete example, we start by introducing briefly the HVDM model in Section~\ref{model}. In Section~\ref{sec:SS} we present the particularities of the self-interaction processes, with or without light mediators.
In section~\ref{sec:DMrd} we  introduce the various ways in which the DM relic density can be produced and draw some general statements with regard to which ones could be compatible with large self-interactions.
In Section~\ref{sec:NoLightMediator}, we consider one  by one and in detail the various production regimes for the HVDM setup in the case where there is no light-mediator
enhancing the self-interaction cross section. We explain why we do not find any viable situation with the 2-to-2 annihilations, and subsequently consider the 3-to-2 possibility.
In Section~\ref{sec:LightMediator} we proceed in a similar way for the case of a light mediator enhancing DM self-interactions.
Finally, we present our conclusions in Section~\ref{conclusions}.

\section{Hidden Vector Dark Matter: Description of the Model}
\label{model}

Here we provide the essential ingredients of the HVDM model~\cite{Hambye:2008bq}, which is the one we will consider explicitly in all subsequent sections. This model is defined by the simple Lagrangian
\begin{equation}
{\cal L}= {\cal L}_{SM} -\frac{1}{4} F'^{\mu\nu} \cdot F'_{\mu \nu}
+(D_\mu \phi)^\dagger (D^\mu \phi)
-\mu^2_\phi\,\phi^\dagger \phi -\lambda_\phi\,(\phi^\dagger \phi)^2 -\lambda_m\,\phi^\dagger\phi\,H^\dagger H\,,
\label{Lag0}
\end{equation}
where ${\cal L}_{SM}$ is the SM Lagrangian. The $F'_{\mu\nu}$ field strength tensor refers to a new $SU(2)_X$ gauge symmetry, with gauge coupling $g_X$ and gauge bosons $A'_\mu$, under which all the SM particles are singlets. $\phi$ is a $SU(2)_X$ scalar doublet which carries no SM charges.
Its $vev$, $v_\phi$, breaks $SU(2)_X$ spontaneously, which leads to a single real scalar, $\eta'$, and three massive gauge bosons, $A^i_\mu$~$(i=1,2,3)$. 
Due to a remnant accidental custodial $SO(3)$ symmetry, these three gauge bosons form a triplet with the same mass $\ma=g_X v_\phi/2$,  and are therefore automatically stable. These are the DM candidates of the model. Such a gauge structure is allowed to communicate with the SM through a unique renormalizable interaction, $\lambda_m$, of the Higgs portal type.

In this setup the $\eta'$ scalar mixes with the SM scalar $h'$, which, after diagonalization, leads to the physical $\eta$ and $h$ fields. The corresponding masses are
\begin{equation}
\mh^2 = 2\lambda v^2 c_\beta^2 + 2\lambda_\phi v_\phi^2\,s^2_\beta - \lambda_m v v_\phi\,s_{2\beta}\,,~~~~\meta^2 = 2\lambda_\phi v_\phi^2\,c_\beta^2 + 2\lambda v^2 s^2_\beta + \lambda_m v v_\phi\,s_{2\beta}\,,
\end{equation} 
where $s_\beta\equiv\sin\beta$, $c_\beta\equiv\cos\beta$, with the mixing angle given by {$\tan2\beta = \lambda_m v v_\phi/(\lambda_\phi v_\phi^2 - \lambda v^2)$. 
Here $v=246$\,GeV and $\lambda$ is the self-coupling of the SM scalar doublet.
For $m_\eta \ll m_h$, this gives the simple relation, $|\lambda_m| \simeq (g_X\sin2\beta \cdot m_h^2)/(4m_Av)$.}
Notice that the model is based on only four independent parameters, which can be taken, for example, to be $\ma$, $\meta$, $\sin\beta$ (or $\lambda_m$)  and $g_X$ (or $\alpha_X \equiv g_X^2/4\pi$).

\begin{figure}[t]
\begin{center}
\includegraphics[width=0.3\textwidth]{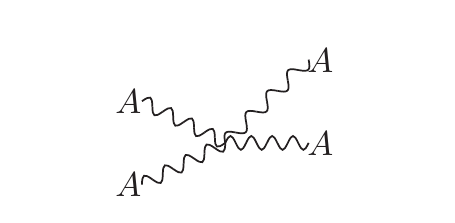}
\includegraphics[width=0.3\textwidth]{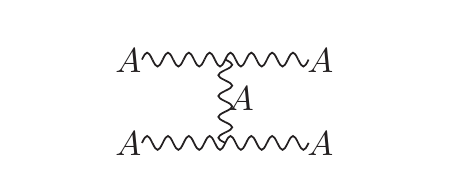}
\includegraphics[width=0.3\textwidth]{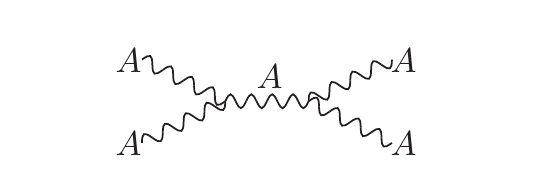}\\
\includegraphics[width=0.3\textwidth]{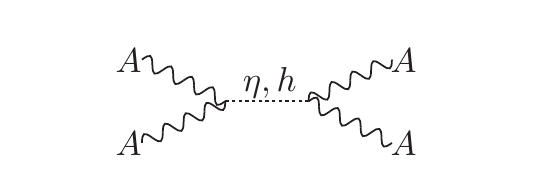}
\includegraphics[width=0.3\textwidth]{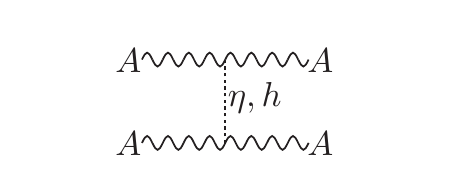}
\end{center}
\caption{ \textit{Diagrams for dark matter self-interactions.}
}
\label{fig:selfDM}
\end{figure}

\begin{figure}[t]
\centering\includegraphics[width=0.8\textwidth]{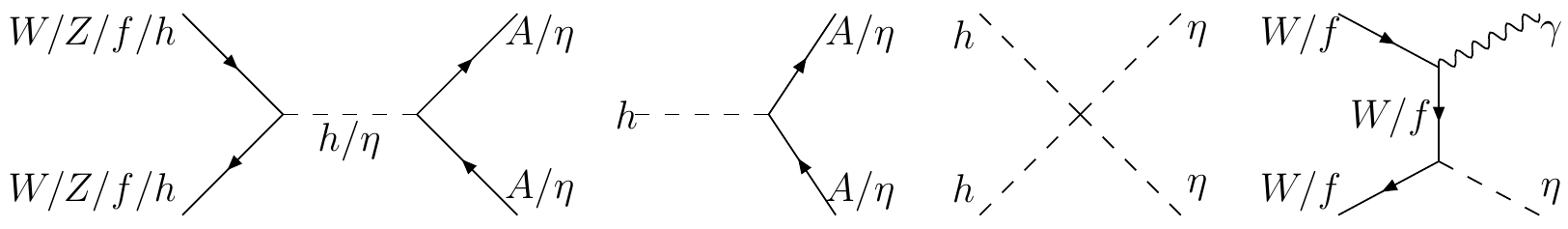}
 \caption{Most relevant processes that transfer energy through the Higgs portal mixing between  the visible sector (i.e.~SM particles) and the HS sector ($\eta$ and $A$ particles).}
 \label{SMtoDM:channels}
\end{figure}

In Appendix~\ref{sec:Higgsdec} we show current constraints on the mixing angle $\beta$ due to invisible decay of the SM scalar boson $h$. In general, they imply small mixing angles. As a result, one can distinguish between a HS, consisting of the DM fields and the scalar $\eta$,  and the visible sector consisting of the SM particles, both of which are connected via the $\lambda_m$ connector interaction. For small $\sin \beta$, the DM particles have the following interactions 
\begin{eqnarray}
{\cal L}_{\rm int} &=& {\cal L}_{HS} + {\cal L}_{connect} \label{Lag1}\,,\\
{\cal L}_{HS} &=& \df{g^2_X}{8}\left[ 4\, \partial_\mu A_{\nu} \,A^{\mu} A^{\nu}
- 2\, (A_\mu A^{\mu})^2+ A_{\mu} A^\mu\, \eta^2 + 2 \,v_\phi\, A_{\mu}A^\mu \,\eta \right]\,,\\
{\cal L}_{connect} &=&  \df{g^2_X}{8}A_{\mu} A^\mu \left( s^2_\beta\,h^2 - s_{2\beta}\,\eta\,h - 2\,s_\beta \,v_\phi \,h\right)
\end{eqnarray}
(where $SU(2)_X$ indices sum is implicit). Here the label $HS$ and $connect$ refer to the interactions within the HS and to the portal interactions between the visible and hidden sectors, respectively. 
Associated to the non-Abelian character of the gauge symmetry, ${\cal L}_{HS}$ includes a trilinear gauge term which allows processes with an odd number of gauge fields. Regarding the purpose of this work, Eq.~\eqref{Lag1} describes the main ingredients of the HVDM model, including DM self-interactions (Fig.~\ref{fig:selfDM}), HS-SM interactions (Fig.~\ref{SMtoDM:channels}) and DM annihilation processes.

\begin{figure}[ht!]
\centering
\includegraphics[width=0.3\textwidth]{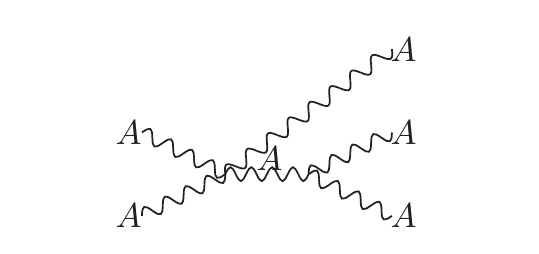}
\includegraphics[width=0.3\textwidth]{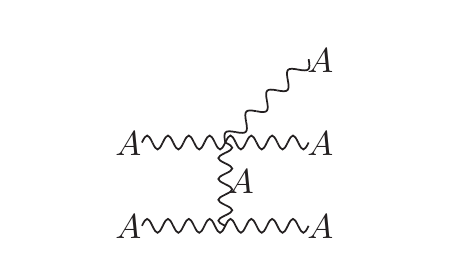}
\includegraphics[width=0.3\textwidth]{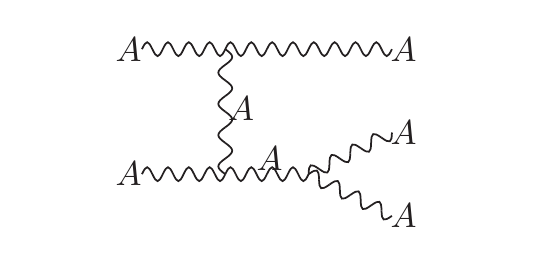}\\
\includegraphics[width=0.3\textwidth]{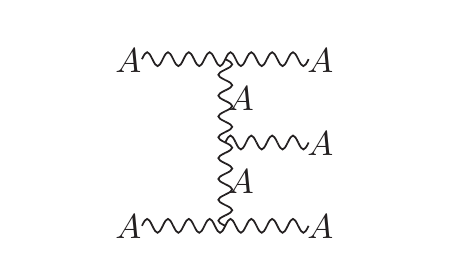}
\includegraphics[width=0.3\textwidth]{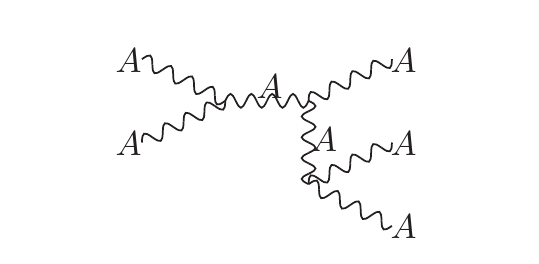}
\includegraphics[width=0.3\textwidth]{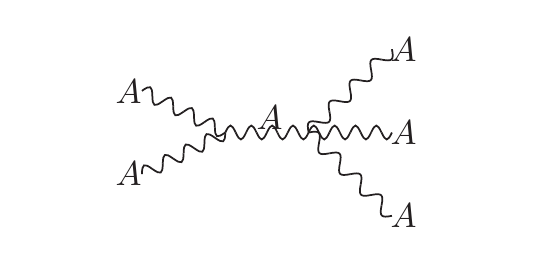}\\
\includegraphics[width=0.3\textwidth]{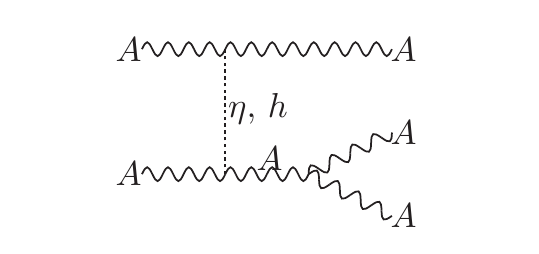}
\includegraphics[width=0.3\textwidth]{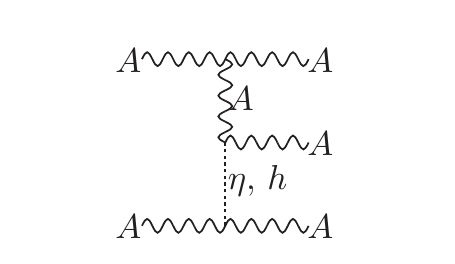}
\includegraphics[width=0.3\textwidth]{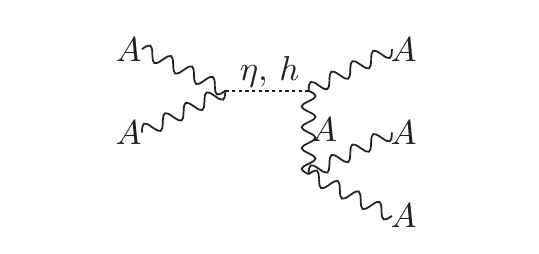}\\
\includegraphics[width=0.3\textwidth]{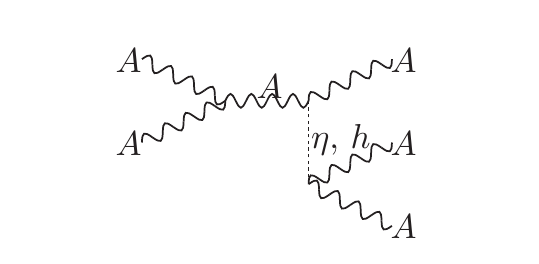}
\includegraphics[width=0.3\textwidth]{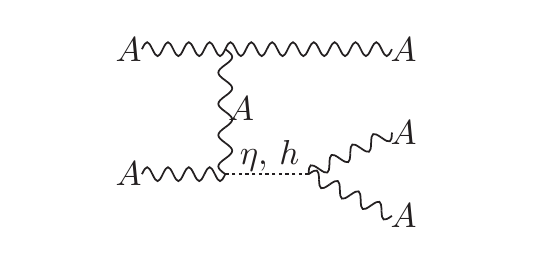}
\caption{ \textit{Tree level diagrams for the process $AA\leftrightarrow AAA$.}
}
\label{xx-xxx}
\end{figure}

As is well known, 3-to-2 processes are forbidden in most well-established DM setups, where the DM stability is guaranteed by a $\mathbb{Z}_2$ symmetry.\footnote{In $\mathbb{Z}_2$ symmetric DM models, however, 4-to-2 annihilations are allowed and could be the dominant ones~\cite{Bernal:2015xba}.}
One possible way to circumvent this issue is to consider a different discrete symmetry such as a $\mathbb{Z}_3$~\cite{Bernal:2015bla,Choi:2015bya}. This procedure is rather ad hoc and it is not necessary in models where the DM stability is not imposed by hand but emerges as a result of the DM dynamics~\cite{Hochberg:2014kqa,Yamanaka:2014pva,Lee:2015gsa,Hansen:2015yaa}. HVDM model falls into this category. In fact, in this model, while DM is stable because of the $SO(3)$ accidental symmetry, still the annihilation of three DM particles into two takes place naturally, induced by the gauge nature of the trilinear term. 
For instance, in Fig.~\ref{xx-xxx} we show the tree level diagrams leading to the conversion of three DM particles into two of them.

For the phenomenological study of the following sections, this model is implemented in {\tt FeynRules}~\cite{Christensen:2008py,Christensen:2009jx} and the output in obtained in {\tt CalcHEP}~\cite{Belyaev:2012qa}.


\section{Dark Matter Self-Interactions}
\label{sec:SS}

There are many physical mechanisms that can lead to self-interacting DM. In the HVDM model, this arises in two different ways: via the exchange of the scalar mediators $\eta$ or $h$ (Fig.~\ref{fig:selfDM}-bottom) or via the quartic and cubic hidden gauge interactions (Fig.~\ref{fig:selfDM}-top). If the mediators are heavier than the DM, the latter processes typically dominate. In this case 
the perturbative calculation is reliable and the corresponding cross section depends only mildly on the non-relativistic DM velocity. In particular, in the limit of small mixing and vanishing relative velocity, the Born approximation cross section reads 
\begin{equation}
\label{sigmaAAAApert}
\sigma_{AA}=\frac{65}{9}\,\frac{\alpha_X^2}{\ma^2} \left[1+ {\cal O}\left(\frac{\ma^2}{\meta^2} \right)\right]\,.
\end{equation}
Of particular importance is the fact that in this case, if $\alpha_X\lesssim 1$, in order to obtain the order of magnitude associated to the solution of small-scale problems, i.e. $\sigma_{AA}/\ma\sim 0.1$~cm$^2$/g, one needs $\ma \lesssim 400$\,MeV.

On the other hand, when the scalar mediators are significantly lighter that the DM, non-perturbative effects may become important.
In the HVDM model, for small mixing angles  (so that Higgs-mediated processes are negligible), these effects are encoded in the attractive Yukawa potential\footnote{Note that there is a prefactor difference with respect to the canonical Yukawa potential.  This results from: $a)$ the way the couplings are normalized, and $b)$ the fact that in our case the colliding particles are vector bosons and not fermions.}
\begin{equation} 
V(r) = -\df{\alpha_X }{16}\df{e^{-\meta\,r}}{r}\,.
\label{pote}
\end{equation} 

Non-perturbative effects become relevant when the range of the potential $1/\meta$ is larger than the `Bohr radius' of a DM pair, $(\alpha_X\,\ma)^{-1}$. Along this work we will refer to this condition, i.e. $m_\eta \lesssim \alpha_X m_A$, as the light mediator case. 
For non-relativistic DM, the self-interacting cross section can be obtained by solving the corresponding Schr\"odinger equation to this potential~\cite{Gould:2006th, Buckley:2009in}. Note that the $t$-channel exchange of scalars displays a collinear divergence that translates into a divergent total self-interacting cross section. However, this is irrelevant for the DM distribution in galaxies because  it is only important when $\cos \theta \rightarrow 1$, which leaves the corresponding DM trajectories unchanged.  Due to this fact, for a light mediator it is more convenient to use the {\it transverse} cross section, $\sigma_T\equiv\int d\Omega\,(1-\cos\theta)\,d\sigma_{AA}/d\Omega$.

Also, for light mediators, a non-trivial dependence on the velocity emerges as a result of the non-perturbative effects. This is crucial because, due to this, the  self-interaction cross section changes, according to the astrophysical system under consideration. DM typical velocities today can vary from $\sim 10$~km/s (in small-scale objects such as dwarf galaxies) to $\sim 1000$~km/s (in clusters). One can qualitatively identify two different regimes depending on the value of the velocity or, more precisely, depending on how the de Broglie wavelength of the DM, $(\ma v)^{-1}$, compares to the range of the potential $1/\meta$. 

The first  regime holds when $\ma\,v/\meta \lesssim 1$. It is known as resonant regime because in this case the formation of quasi-bound states becomes possible and there is an enhancement on the self-scattering cross section by a few orders of magnitude \cite{Buckley:2009in,Tulin:2013teo}. In this work, instead of solving the Schr\"odinger equation, we closely follow Ref.~\cite{Tulin:2013teo}, where it was shown that in this case the Yukawa potential can be well approximated by a Hulthén potential, whose corresponding Schrödinger equation admits an analytical solution, from which the self-interaction cross section can be calculated in the $s$-wave limit. 

On the other hand,  if the range of the potential is much larger than the de Broglie wavelength, i.e. $\ma\,v/\meta \gg 1$, the classical  regime holds. Here, although the perturbative result does not apply, no resonances exist. For this scenario, in order to calculate the self-interacting cross sections, we use the fit reported in Refs.~\cite{Khrapak:2003, Feng:2009hw}, as also summarized in Ref.~\cite{Tulin:2013teo}. 
 
In contrast to the perturbative case, in both the resonant and classical regimes, since the self-interacting cross section is enhanced by non-perturbative physics,  the DM mass does not have to necessarily lie below a few hundreds of MeV.

\vspace{.1cm}

DM self-interactions induced by a heavy mediator, denoted as $\sigma_{AA}$, or by a light mediator, denoted as $\sigma_T$, would lead to slightly different effects in observations \cite{Kahlhoefer:2013dca}. Nevertheless, for the sake of simplicity, we will not distinguish them any further in this work. Instead, we assume the same bounds on both, and take $\sigma_{AA} (\sigma_{T})/m_A \lesssim 1$\, cm$^2$/g as derived from observations of colliding clusters ($v \sim 1000$~km/s). Such a simplification is in practice justified, given the large uncertainties in existing astrophysical data.  At last, in order to address the small scale problems in the HVDM model, we only focus on self-scattering cross section in the range $0.1\lesssim\sigma_{AA}(\sigma_{T})/\ma\lesssim10$ cm$^2/$g on dwarf galaxy scales ($v \sim 10$~km/s), and refer to this condition as \emph{self-interaction hypothesis} throughout the paper.

\section{Various Possibilities of Accounting for the Dark Matter Relic Density}
\label{sec:DMrd}

If DM belongs to a HS which communicates to the SM sector through a portal, as is the case for the HVDM model, there are several ways  along which the DM relic density could have been produced. From the less coupled case to the most coupled case, and following Ref.~\cite{Chu:2011be}, the main ways are:
\begin{itemize}

\item[0)] \underline{\it No portal.} The level zero is to consider no connector at all or a very tiny one, such that it has no effect on the DM relic density. In this case the HS has to be created independently of the visible sector at the end of inflation and it has a history which can be probed only through DM gravitational and self-interacting  effects.

\item[1)] \underline{\it Freeze-in.} The next possibility is based on both {\it (i)} a DM particle which does not thermalize in the HS (i.e.~tiny $\alpha_X$) and {\it (ii)} a portal  which is small enough for both sectors not to thermalize, but large enough to create DM particles out of the visible sector (i.e.~through the $\lambda_m$ interactions). 
This is the \emph{freeze-in} mechanism~\cite{Moroi:1993mb,McDonald:2001vt,Hall:2009bx}, where the final DM abundance is uniquely determined by the strength of the portal interactions, such as the processes illustrated in Fig.~\ref{SMtoDM:channels}. 
This regime does not need large values of the portal, typically requiring $\lambda_m\sim {\cal O}(10^{-12})$ to ${\cal O}(10^{-10})$, due to the fact that, unlike in the usual freeze-out scenario, there is no Boltzmann suppression for the DM density. The necessary suppression of the production rate precisely comes from the smallness of the portal. An assumption typically made along this regime is that after reheating the number of DM particles was negligible with respect to the number created through the portal. {The bulk of the relic density will be created through the mediator and determined solely by the mass and interactions of the particles involved in the model.}

Clearly this scenario  generically does not apply to the case where the required amount of self-interactions proceed through a heavy mediator because. If that is the case, the self-interaction hypothesis
leads to either a large self-interacting coupling ($\alpha_X$) and thus to the thermalization of the HS with itself or to a very low $\ma$ at the verge of hot DM, as we will see in Section~\ref{sec:NoLightMediator}. For the light mediator case nevertheless, we will show in Section~\ref{sec:LightMediator} that the freeze-in scenario is interestingly viable, due to the fact that in this case  the self-interaction hypothesis does not require neither so large values for the HS self-coupling $\alpha_X$  nor a tiny DM mass.

\item[2)] \underline{\it Reannihilation.}
For same values of the portal interaction than in the freeze-in regime, if one increases the HS interaction, still both sectors do not thermalize with each other but, at some point, enough HS particles have been created through the portal and the HS interactions are strong enough to produce a dark thermal bath. 
The  condition for the non-thermalization of both sectors with each other is $n_A^\text{eq}(T)\,\sqrt{\langle \sigma v \rangle_{HS}\,\langle \sigma v \rangle_{connect}}\lesssim H$~\cite{Chu:2011be}, where
the \emph{HS} and \emph{connect} labels refer to the HS cross section (from $\alpha_X$) and portal cross section (from $\lambda_m$) respectively, with $n_A^\text{eq}$ the total equilibrium DM number density and $H$ the Hubble expansion rate.
Once the HS has thermalized, the Universe is made out of two thermal baths, each one with its own temperature, $T'$ for the HS and $T$ for the visible sector (photon temperature).
The relation $T'<T$ naturally holds if reheating has occurred mostly in the visible sector as we will assume here.

 Here, as in the freeze-in regime, both sectors communicate only through out-of-equilibrium creation of HS particles from the visible sector through the portal, Fig.~\ref{SMtoDM:channels}.
This leads to a situation where once $T'$ goes below $\ma$ the HS interactions get Boltzmann suppressed and freeze. 
The dominant annihilation channels, which diminish the abundance of $A$, are DM (semi-) annihilations into $\eta$ scalars for $\ma>\meta$, or $AAA\rightarrow AA$ for $\ma \ll \meta$.  Nonetheless, 
if the portal interaction still produces HS particles when the equilibrium number density gets largely Boltzmann suppressed (at $T'\sim\ma/20$), the final DM number density does not freeze  in the same way as in the ordinary freeze-out mechanism. In this case one lies in the so-called reannihilation regime, where the DM freezes later, after a period of quasi-static equilibrium between the source term which creates DM particles and the HS processes which annihilate them (see Refs.~\cite{Cheung:2010gj,Cheung:2010gk} for a decay process and Ref.~\cite{Chu:2011be} for a scattering process).

We will see that for the HVDM setup with no light mediator to enhance the self-interaction, this regime does not apply because in this model the source term has already stopped to be active before DM freezes out in the HS. 
The discussion of this regime in presence of a light mediator is more involved as we will see in Section~\ref{sec:LightMediator}.

\item[3A)] \underline{\it Dark freeze-out with $T'<T$.} 
As just mentioned, with $T'<T$, if the source term is still active at the time when the DM annihilations become largely Boltzmann suppressed,
one lies in the reannihilation regime. If instead, still with $T'<T$, the source term has already ceased to be active at this time, one has a freeze-out process in the HS (`dark freeze-out')~\cite{Feng:2008mu,Ackerman:mha}. It is analogous to the usual freeze-out except for the fact that it occurs in a 
separated sector with a temperature $T'$ smaller than $T$. 
In Section~\ref{subsec:darkFOnoequil} we will show in detail that this regime can be easily successful for the case of no light mediator to enhance the self-interactions, if the freeze-out is dominated by 3-to-2 processes.

For a 2-to-2 freeze-out with a light mediator, the situation is completely analogous to the reannihilation case, as we will see in Section~\ref{sec:LightMediator}.

\item[3B)] \underline{\it Dark freeze-out with $T'=T$.} 
Increasing even more the connector interaction, for example for $\lambda_m\gtrsim 10^{-7}$ if $m_A=10$\,GeV, the two sectors thermalize ($T'=T$)
and the relic density results from an ordinary freeze-out driven by the HS interaction.
We will show in Section~\ref{subsec:darkFOequil} that this regime does not work for a heavy mediator under the assumption of perturbativity. That is, the dark coupling needed to obtain both the observed relic abundance and self-interactions is too large. Also, the kinetic equilibrium between the HS and the visible sector in this model can only be established from assuming extra non-renormalizable interactions.  Moreover, in the case of a light mediator this regime is in strong tension with various experimental constraints.

\item[4)] \underline{\it Portal interaction freeze-out.}
By increasing even more the connector interaction, one ends up with the last regime, where not only both sectors thermalize as in the previous case, but also the freeze-out process is dominated by connector processes, rather than by HS processes. As we will see, this regime does not  work under the self-interaction hypothesis anyway.

\end{itemize}

\begin{figure}[t]
\centering
\includegraphics[width=0.47\textwidth]{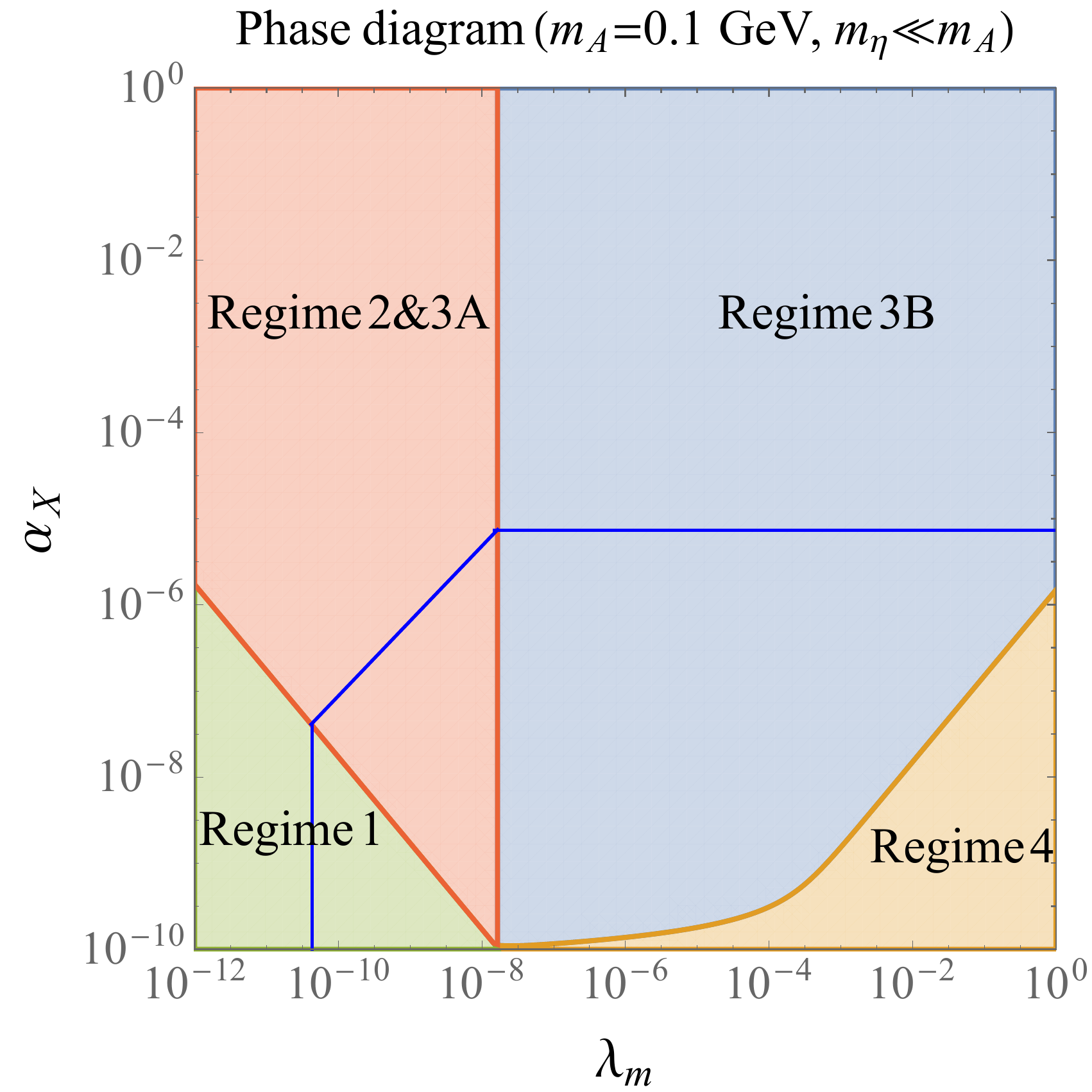} 
\includegraphics[width=0.47\textwidth]{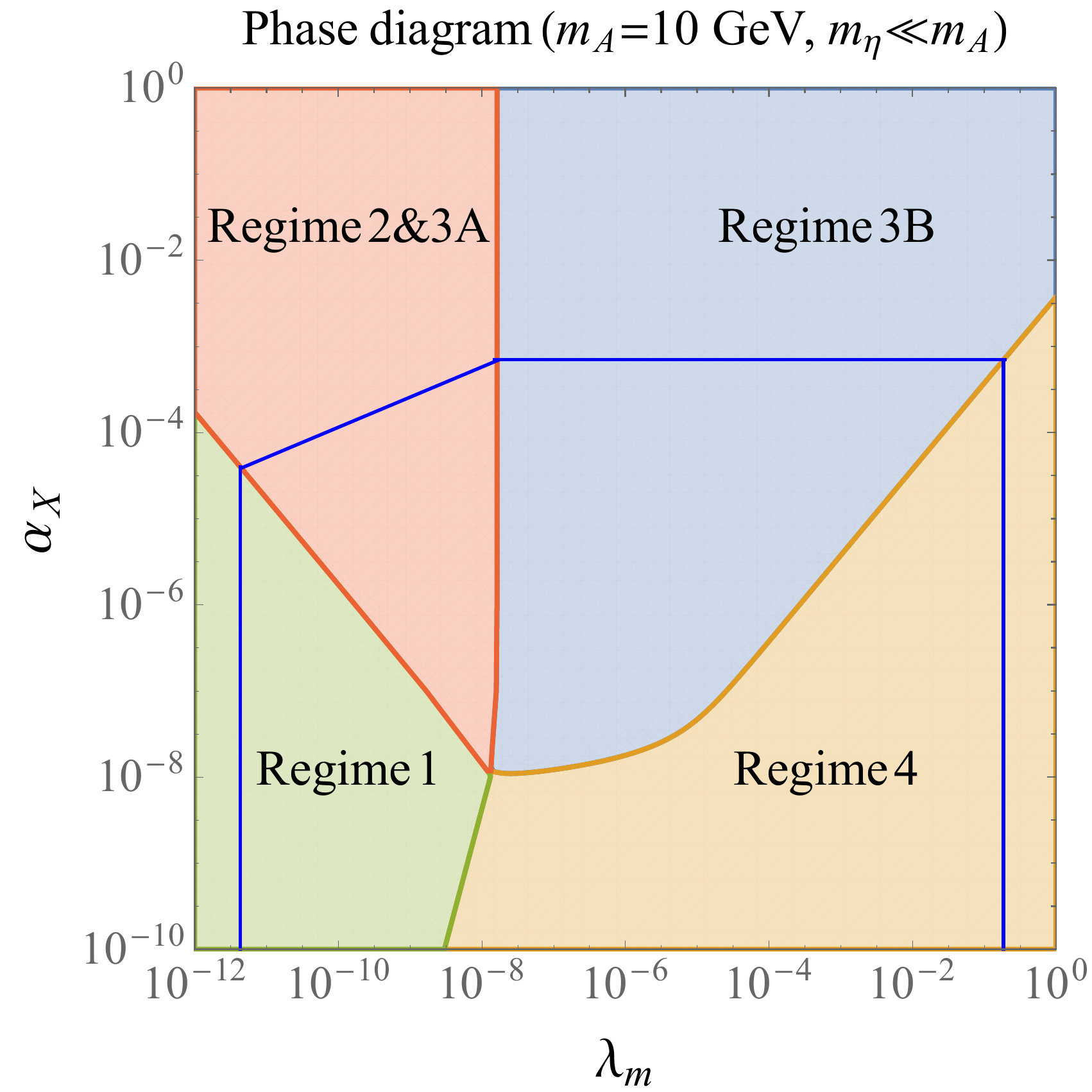} 
\caption{Phase diagrams for $m_{A}=0.1$ and 10\,GeV, in the case $m_{\eta}\ll\ma$, assuming that the HS has been created through the portal. The solid blue line shows the values of $\lambda_m$ and $\alpha_X$ which lead to the measured DM relic density. Above (below) this line one gets a too small (large) relic density. Due to the high computational time required for analyzing regimes 2 and 3A, we do not try here to separate them.
}
\label{phasediagram}
\end{figure}

For the HVDM setup, plotting which regime corresponds to a given set of parameters is not possible
in general because it depends in a complex way on four independent parameters. However for the light mediator case, $m_\eta\ll m_A$, the dependence of the phenomenology on $m_\eta$ is tiny, and hence for a fixed value of $m_A$ one can show the various regimes according to the values of $\alpha_X$ and $\lambda_m$. As an example, in Fig.~\ref{phasediagram} we schematically show the phase diagram for $\ma=0.1$ and $10$\,GeV, as will be explained in Section~\ref{sec:LightMediator}.  The parameter space giving rise to the observed relic density is sketched as a solid blue line.

In the next Sections we will consider each regime one-by-one in more detail, in particular the two regimes which can easily fulfill all constraints: dark freeze-out via 3-to-2 processes with $T'<T$ and freeze-in with a light mediator. For each case, we will determine the parameter space of the model which addresses the small-scale problems and is in agreement with the relic density, DM direct and indirect detection, BBN, CMB, $X$-ray constraints, as well as the cluster bound on self-interactions.


\section{Scenario with no Light Mediator
}
\label{sec:NoLightMediator}

In this Section we consider the case where the self-interactions are not enhanced by any light mediator.
For the HVDM model this means that $\meta$ is larger than $\ma$.\footnote{Strictly speaking, in this scenario one can also consider the case when $\meta$ is slightly lighter than $\ma$, because in that situation there is no enhancement of the self-interacting cross section due to quantum mechanical effects associated to the lightness of the mediator.
Nevertheless, if $m_\eta\lesssim \ma$, 2-to-2 DM annihilation to $\eta$ scalars is the dominant process and the scenario is not viable.
One encounters the usual problem of 2-to-2 frameworks without light mediator to enhance the self-interaction: large self-interactions (driven by $\alpha_X$) imply a very fast annihilation rate (also driven by $\alpha_X$), i.e.~a very suppressed relic density.}
Importantly, as discussed in Section~\ref{sec:SS}, in this case large enough self-interactions require $\ma\lesssim 400$\,MeV and a large value of the associated coupling $\alpha_X$. 
Moreover, such a light DM also requires a very suppressed portal in order to avoid observable invisible decay of the SM  scalar (as given by Eq.~\eqref{invi:Higgs} of Appendix~\ref{sec:Higgsdec}).
With this in mind, we consider now the various DM production regimes introduced in Section~\ref{sec:DMrd}. For convenience we discuss them in the reverse order, i.e.~we will consider them from the usual cases where both sectors are the most coupled to the less considered cases where they are less coupled. This Section illustrates the observation that in simple models with no light mediator to enhance the self-interaction cross section, it is  in general difficult to accommodate all constraints, unless one considers a freeze-out DM production regime with 3-to-2  annihilations and $T'<T$. 

\subsection*{Regime 4: Portal Interaction Freeze-out}
\label{subsec:connectorFO}

This regime corresponds to the usual thermal freeze-out, where a couple of DM particles annihilate into a pair of SM particles. In the HVDM setup it is not viable.
In fact, for $\ma\lesssim 400$\,MeV, the only kinematically allowed final states consist of light fermion pairs, with a corresponding thermally averaged annihilation cross section $\langle\sigma v\rangle_{AA\to f\bar f}\sim{y_f^2\,\alpha_X\,\sin^2\beta}/{\ma^2}$.
Such cross section is very suppressed because of the small mixing angle and the Yukawa coupling $y_f$, far from the canonical value of few $10^{-26}$~cm$^3$/s~\cite{Steigman:2012nb}. Thus it would give rise to an overclosed Universe.
In addition, if one assumes values of $\alpha_X$ large enough to fulfill the self-interaction  hypothesis, the connector processes are slower than the HS ones, so that the DM production is dominated anyway by HS processes.
This brings us to the next regime.

\subsection*{Regime 3B: Dark Freeze-out with $T'=T$} 
\label{subsec:darkFOequil}

As explained in Section~\ref{sec:DMrd}, the next most coupled regime assumes a connector (from $\lambda_m$) sufficiently large for both sectors to reach kinetic equilibrium with each other, $T=T'$, but small enough for DM annihilation into SM particles to be slower than DM annihilations into HS particles.
If $m_\eta> m_A$, as we consider here, the $AA\to \eta\eta$ and $AA\to A\eta$ annihilation channels are kinematically closed. In this case only the 3-to-2 processes and connector processes are possible.
It turns out that for the part of the parameter space which is relevant for self-interactions (i.e. large enough $\alpha_X$ and $m_{A}\lesssim 400$\,MeV), even if the 3-to-2 processes are phase suppressed they are still faster than the (Yukawa suppressed) connector processes. Therefore they dominate the DM production.

Concretely, the 3-to-2 interaction rate is given by $ \Gamma_{3\to2} = n_\text{eq}^2 \langle\sigma v^2\rangle_{3\to2}$, where $n_\text{eq}$ is the DM equilibrium number density and $\langle\sigma v^2\rangle_{3\to2}$ is the generalized thermal averaged cross section which, from dimensional arguments, scales as $\alpha_X^3/\ma^5$~\cite{Hochberg:2014dra}.  
This gives a rate ratio
\begin{equation}
\frac{\Gamma_{AA\to f\bar f}}{\Gamma_{3\to2}} =\frac{ 
n_\text{eq} \langle\sigma v\rangle_{AA\to f\bar f}}{n_\text{eq}^2 \langle\sigma v^2\rangle_{3\to2}} \sim
\frac{1}{9\, \left(\df{\ma T'}{2\pi}\right)^{3/2} e^{-\ma/T'}}
\frac{\df{y_f^2\,\alpha_X\sin^2\beta}{\ma^2}}{\df{\alpha_X^3}{\ma^5}}
\sim \df{y_f^2\,\sin^2\beta}{10^{-10}\,\alpha_X^2} \ll 1\,,
\label{Gamma22}
\end{equation}
where in the next-to-last step we assume a ratio of DM mass to dark temperature $T'$ close to 20 (cf. Eq.~\eqref{xpf} in the Appendix). 
In other words, a scenario whose mediator is heavier than the DM particle and at the same time compatible with the solution of the small-scale problems predicts that the dominant annihilation mechanism is a 3-to-2 annihilation (this was actually one of the original motivations in Ref.~\cite{Hochberg:2014dra}).

Note that since the $\eta$ boson can not be arbitrarily heavy with respect to the DM,\footnote{
Both particles acquire their masses by the breaking of the hidden gauge symmetry $SU(2)_X$.
In fact, the perturbativity of the scalar potential implies
\begin{equation}
\lambda_\phi = \left(\frac{m_h^2 \sin^2\beta + \meta^2 \cos^2\beta}{8\ma^2}\right) g_X^2 \approx \left( \frac{ \meta}{\ma} \right)^2 \left(\frac{\pi\,\alpha_X}{2} \right) \lesssim 4 \pi \,,
\end{equation}
which leads to $\meta/\ma < \sqrt{8/\alpha_X}$, and hence to a $\eta$ boson only modestly heavier than the DM.
} in principle not only the $AAA\to AA$ annihilation must be considered but also 3-to-2 processes involving $\eta$ scalars in the initial and final states. However, in practice it turns out that the latter are subdominant due to the following reasons. On the one hand, at the time of DM freeze-out, when the 3-to-2 process stops taking place, the $\eta$ particles have effectively disappeared from the thermal bath because they either decay into DM particles or (co-)annihilate very efficiently into DM pairs via 2-to-2 processes. On the other hand, calculating analytically  the generalized cross sections $\langle\sigma v^2\rangle$ in the non-relativistic limit for the processes $AAA\to AA$, $AAA\to A\eta$ and $AAA\to \eta\eta$, we find that 
$\langle\sigma {v^2}\rangle_{AAA\to AA}/\langle\sigma {v^2}\rangle_{AAA\to A\eta}\sim 10$ and $\langle\sigma {v^2}\rangle_{AAA\to AA}/\langle\sigma {v^2}\rangle_{AAA\to\eta\eta} \sim 17$. As a result, it is safe to neglect the channels involving $\eta$ without loss of generality and consider only the process $AAA\to AA$. In this way, this setup is a realization of the scenario studied in Refs.~\cite{Carlson:1992fn, Hochberg:2014dra}. The cross section $\langle\sigma {v^2}\rangle_{AAA\to AA}$ depends mostly on $\ma$ and $\alpha_X$, provided that the mixing angle is small, $\sin \beta\ll 1$. There is also an additional dependence on $\meta$, but this is negligible as long as $1 < \meta/\ma < \sqrt{8/\alpha_X}$.

\begin{figure}[t!]
\centering
\includegraphics[width=0.49\textwidth]{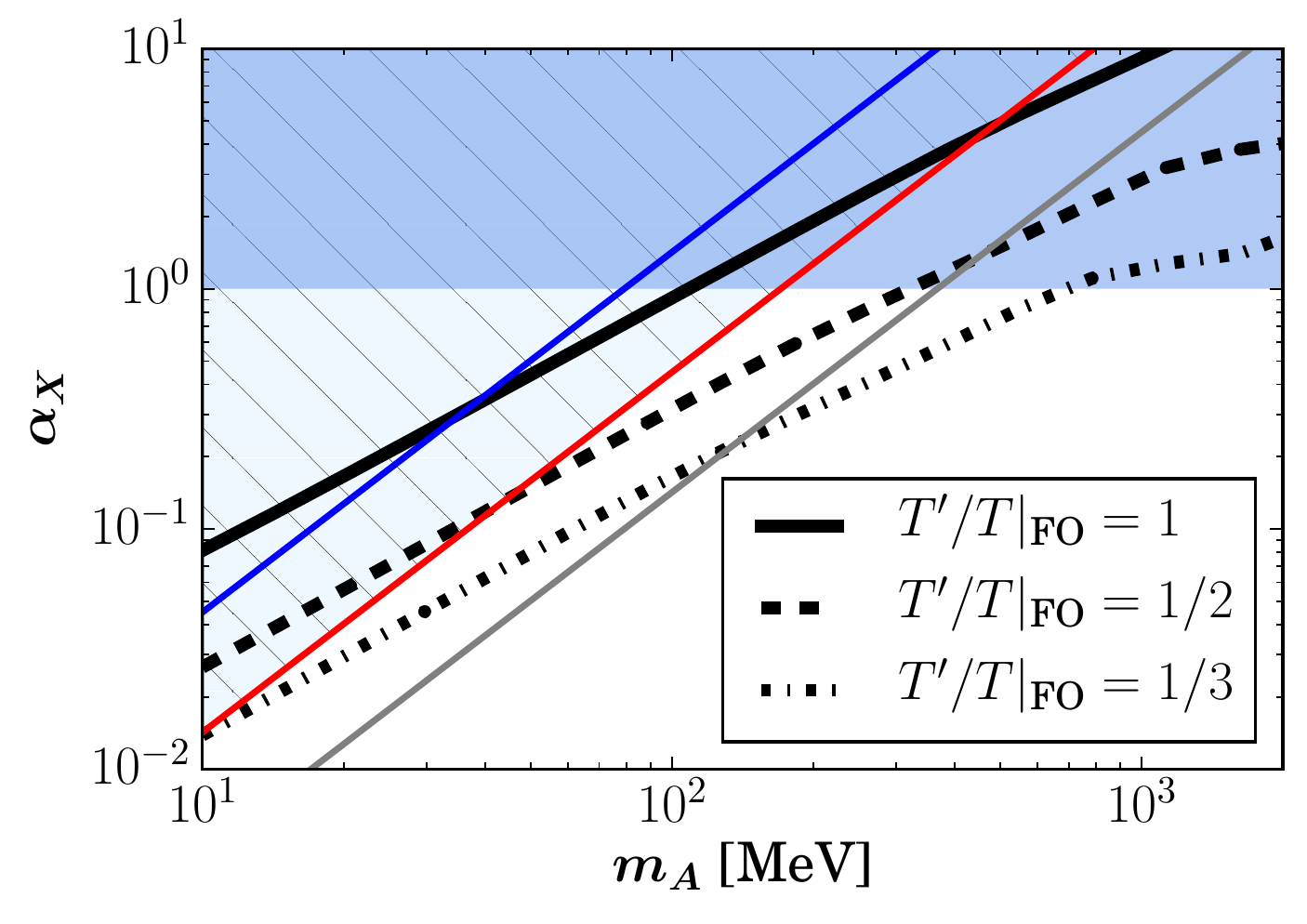} 
\includegraphics[width=0.49\textwidth]{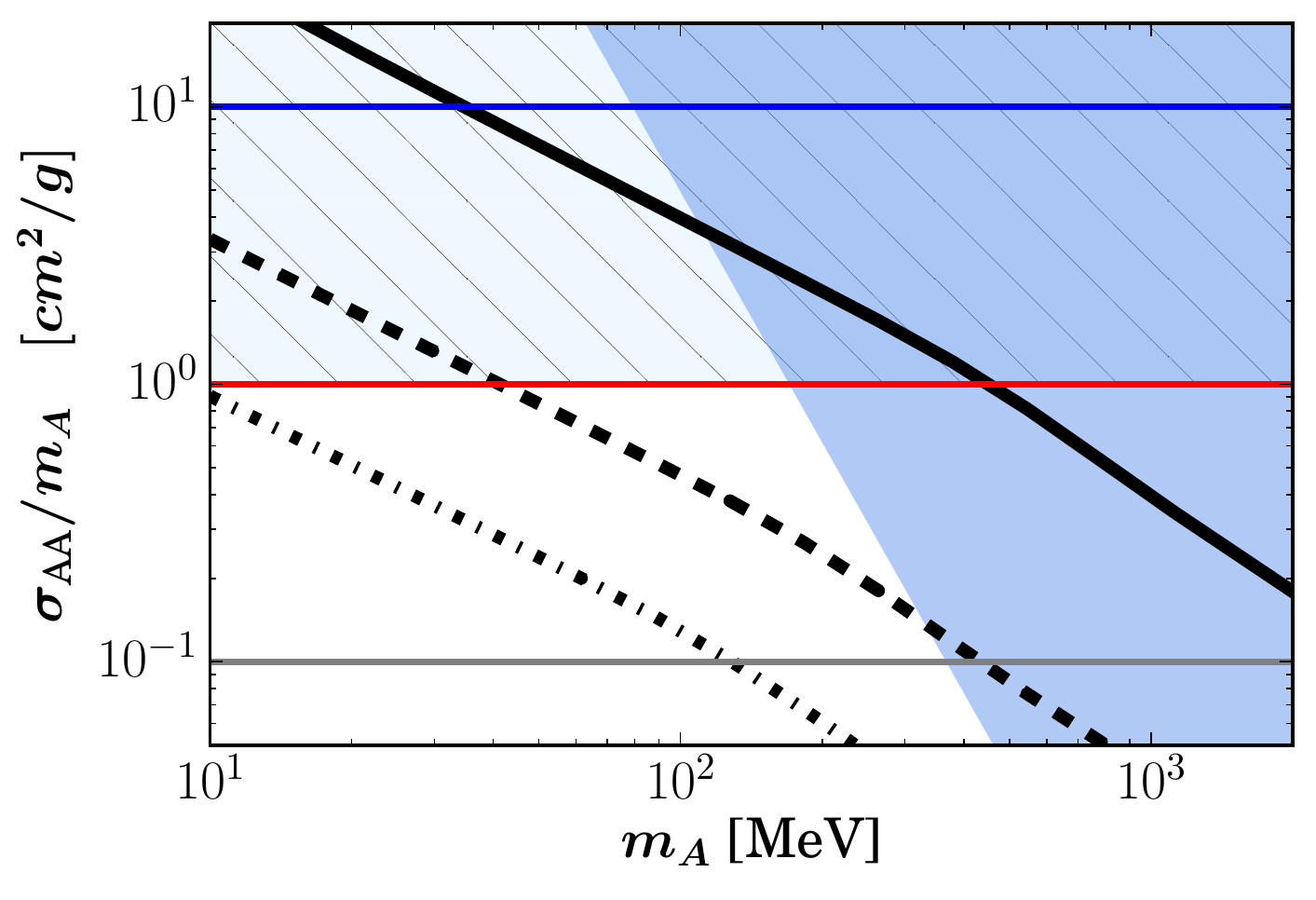} 
\caption{ Values of $\alpha_X$ and $m_A$ which yield the observed DM relic density. The black solid lines correspond to thermal production of DM via the 3-to-2 mechanism under the assumption of kinetic  equilibrium (i.e.~$T=T'$) until freeze-out. 
The gray, red and blue lines  give a value of $\sigma_{AA}/\ma$ equal to 0.1, 1 and 10 cm$^2/$g, respectively.
The area where perturbation theory breaks down, $\alpha_X>1$, is shown  in blue, whereas the region excluded by cluster observations is the hatched one. We also  give the values of $\alpha_X$ and $m_A$ which yield the observed DM relic density assuming $T'/T$ ratios at freeze-out different from one (black dashed lines).} 
\label{ma-alpha-som}
\end{figure}

Assuming that temperatures of both sectors are equal at least until the chemical freeze-out, the evolution of the DM relic density can be solved using the Boltzmann equation given in Appendix~\ref{sec:BEandFO}, Eq.~\eqref{BE}. In particular, we obtain the cross section $\langle \sigma v^2\rangle_{3\to2}$ that matches the observed relic abundance using the instantaneous freeze-out approximation 
\begin{equation}
\langle \sigma v^2 \rangle_{3\to2} \approx \left( 8.65\,\text{GeV}^{-5}\right) \,\xf^4 \, g_{*\text{FO}}^{-1.5} \left(\frac{1\, \text{GeV}}{\ma} \right)^2\,,
\label{FOconditionText}
\end{equation}     
with $x_\text{FO}\equiv m_A/T_\text{FO}$ given by Eq.~\eqref{xpf}.
Using the analytical expression for $\langle\sigma v^2\rangle_{3\to2}$, 
 in the left panel of Fig.~\ref{ma-alpha-som} we show (black thick solid line, labeled as $T'/T|_\text{FO}=1$) the value of gauge coupling that reproduces the observed DM relic abundance as a function of $\ma$, assuming kinetic equilibrium between both sectors.

Using Eq.~\eqref{sigmaAAAApert}, the self-interaction hypothesis  determines a region in the plane $\alpha_X-\ma$ lying between the straight diagonal lines in the left panel of Fig.~\ref{ma-alpha-som}, where gray, red and blue lines correspond to $\sigma_{AA}/\ma=0.1$, 1 and 10 cm$^2/$g, respectively. 
 The region for which the perturbative calculation we have done is not expected to be reliable ($\alpha_X>1$) is shown in blue. The parameter space disfavored by cluster observations (i.e.~where $\sigma_{AA}/\ma>1$~cm$^2$/g) is shown in hatched light-red. In the right panel we present the same information in a slightly different way,  showing on the vertical axis the corresponding values of the self-interacting cross section.

Fig.~\ref{ma-alpha-som} shows that the measured relic abundance can not be obtained for perturbative values of 
$\alpha_X$ unless one allows the area which is disfavored by cluster observations. We therefore conclude that the 3-to-2 mechanism is either in tension with experimental data or requires a more elaborated analysis that includes non-perturbative effects~\cite{Hansen:2015yaa}.\footnote{ For the HVDM model, one can expect  that
in its confined phase \cite{Hambye:2009fg} the scalar bound states are lighter than the vector bound states. In that case, the 2-to-2 annihilations of the DM vector bound states to scalar bound states dominate over the 3-to-2 processes.}

We close this subsection by discussing an additional problem one encounters in this regime related to kinetic equilibrium.
As mentioned in the introduction, an important issue concerning 3-to-2 annihilations comes from the fact that when three DM particles annihilate into two HS particles, a sizable part of the rest energy is transformed into kinetic energy.
This ability of the HS to reheat itself can modify significantly structure formation~\cite{deLaix:1995vi}.
One way to avoid this is to assume kinetic equilibrium between the DM and the SM, at least until the DM freezes-out~\cite{Hochberg:2014dra,Choi:2015bya}, so that the SM particles absorb the released kinetic energy in the 3-to-2 reactions.\footnote{Other possibilities have been considered in the literature, e.g. an enlarged HS with relativistic particles, such as Goldstone bosons, which absorb the energy released in the 3-to-2 reactions~\cite{Bernal:2015bla}. We do not consider these possibilities here.}
This is basically what we have assumed above to derive the results of Fig.~\ref{ma-alpha-som}  (for $T'/T|_\text{FO}=1$). However such an assumption is far from granted as it requires a sizable portal interaction between both sectors.

Since we deal with temperatures around 100\,MeV, kinetic equilibrium could be driven from interactions with electrons. A generic way of describing such a portal interaction below the electroweak scale is to use the effective operator
\begin{equation}\label{lagkineq}
{\cal L} = \frac{\, 1}{\Lambda^3} \left( {\cal D}_\mu \phi^\dagger {\cal D}^\mu \phi \right) \bar{e}{e}
\xrightarrow{\small SU(2)_\text{HS}\to SO(3)}
 \frac{\ma^2}{\Lambda^3} (A_\mu A^\mu)\cdot ( \bar{e}{e})
\,,
\end{equation}
which mediates DM scattering off electrons.  Notice that this operator does not break the custodial symmetry $SO(3)$.
In the HVDM model, such an operator is induced by the exchange of $\eta$ scalars, with $\Lambda\sim [\meta^2\,\ma/(g_\phi\, y_e\, \sin\beta )]^{1/3}$. More generally, such an effective operator can be obtained by the exchange of beyond the SM particle(s) between the electrons and the DM particles.
The elastic scattering cross section induced by  this operator is 
\begin{equation}
\langle \sigma v \rangle_\text{kin} \equiv\frac{\epsilon^2}{\ma^2}\,,\hspace{30pt}\text{with}\hspace{10pt}
\epsilon\approx5.6\times 10^6\left(\frac{ \ma}{\Lambda}\right)^3\,,
\end{equation}
where a dimensionless parameter $\epsilon$ is defined to characterize the strength of the scattering. 

It has been shown that when the annihilation into electrons is slower than 3-to-2 reactions~\cite{Hochberg:2014dra}, as is our case for this regime, in order to keep kinetic equilibrium between the DM and the SM, it is necessary that $10^{-9} \lesssim \epsilon \lesssim 10^{-6} $.
This translates into a bound on the scale of the operator of $200$\,GeV $\lesssim \Lambda \lesssim 2$ TeV for DM masses around $100$\,MeV.

In the HVDM model (with only renormalizable interactions), due to the smallness of the electron Yukawa coupling, the lower bound on $\epsilon$ (i.e~upper bound on $\Lambda$) implies that $\sin\beta \gtrsim 10^{-3}$ in the parameter region of interest. However, for the same parameter region, the LHC measurement of the Higgs invisible decay width has excluded $\sin\beta \gtrsim 10^{-5}$ as explained in Appendix~\ref{sec:Higgsdec}. Therefore, we would have to assume that additional new physics is responsible for the operator in Eq.~\eqref{lagkineq}.

As we will show in the next Section this kinetic equilibrium constraint can be avoided in another way, if we assume that both sectors have never been in thermal equilibrium, with a HS temperature $T'$ smaller than the visible temperature $T$, and with a self-thermalized HS due to large enough dark interactions.

\subsection*{Regime 3A: Dark Freeze-out with $T'<T$}
\label{subsec:darkFOnoequil}

In the previous regime two problems  were found. On the one hand, the mixing through the Higgs portal is not enough to maintain the kinetic equilibrium between both sectors, so new physics must be invoked in order to reach it. On the other hand,  with perturbative values of $\alpha_X$, one can not simultaneously reproduce the correct DM abundance and fulfill the  self-interaction hypothesis. 
Interestingly, both problems can be avoided in the  present regime: dark freeze-out without thermalization of both sectors, $T'<T$.

As discussed in Section~\ref{sec:DMrd}, this regime  requires a relatively large value of the HS interaction so that it thermalizes with itself, but still a small connector,
so that the HS does not thermalize with the visible one. Accordingly both sectors naturally have different temperatures.
In this regime it is also assumed that at a temperature $T'\sim \ma/20$ there is no production of DM particles from the visible sector (that is, no reannihilation), so that a freeze-out of the dark interaction occurs just as in the standard freeze-out except for the fact that it occurs in a HS with $T'<T$. 
In fact in the HVDM model, or more generally in models where the connector is of the Higgs portal type, 
this assumption is automatically fulfilled because, as discussed above, the connector $AA\to f \bar{f}$ processes are suppressed by small Yukawa couplings.

Thus the visible and hidden sectors evolve separately from each other, with different temperatures and entropies long before DM freezes-out.\footnote{For other scenarios with 3-to-2 annihilations and different temperatures of the visible and hidden sectors see e.g. Ref.~\cite{Boddy:2014qxa}.} This is the crucial difference between this case and the previous regime.

Assuming different temperatures in both sectors, we solve the Boltzmann equation using the instantaneous freeze-out approximation for two temperature ratios, $T'/T|_\text{FO}=1/2$ and $1/3$, as described in Appendix~\ref{sec:BEandFO}. The corresponding results are overlaid in Fig.~\ref{ma-alpha-som}.  We observe that the measured relic abundance can be obtained for $\alpha_X$ values well in the perturbative range as long as we assume a freeze-out temperature ratio, $T'/T$, smaller than one.  This stems from the following reason. Since the HS remains colder than the SM, the equilibrium DM abundance is smaller than in the case where both sectors thermalize.
During the freeze-out it is thus not necessary to deplete as much DM as before, so a smaller cross section $\langle\sigma v^2 \rangle_{3\to2}$ leads to the observed DM density. Correspondingly, smaller values of $\alpha_X$ are needed in order to get the correct DM abundance for a given DM mass. As a consequence, the 3-to-2 mechanism in this scenario is no longer in tension with perturbativity and cluster constraints, as can been seen from the two dashed and dashed-dotted black lines in Fig.~\ref{ma-alpha-som}. Furthermore, the smaller $T'/T$ is, the colder DM  is during the formation of the first structures. This allows us to surpass the problems associated to the fact that DM heats up itself via 3-to-2 annihilations. As we will see below, for the value of $T'/T$ needed to satisfy the self-interaction  hypothesis and the relic density constraint, such problems turn out not to be a matter of concern anymore.
 In this way, as DM is colder, establishing the kinetic equilibrium is not required  in this regime to satisfy the structure formation constraints.

We now discuss step by step the thermal history of DM in this scenario, including the physics responsible for the difference in temperatures of both sectors at the freeze-out time. This occurs as follows:
 
\begin{enumerate}[$a.$]
\item[(a)] As just mentioned, the regime we consider here requires that both sectors thermalize separately, with $T'<T$. A simple way to justify such a temperature hierarchy is to assume that, during the reheating time, no or very few HS particles were created but that the HS was slowly populated from the visible sector through the various processes of Fig.~\ref{SMtoDM:channels}.  
For a given value of the connector coupling $\lambda_m$, the energy transferred from the visible sector to the HS at $T\simeq M_\text{EW}$ can be estimated to be \cite{Chu:2011be}
\begin{equation}
\rho'/\rho \propto \lambda_m^2\,{M_\text{Pl}/M_\text{EW}}\,,
\end{equation}
with $\rho'$ and $\rho$ the energy densities stored in the hidden and visible sectors, respectively. This means that a tiny value of the connector is already enough to transfer a large amount of energy.
Once a sufficiently large amount of energy has been transferred, i.e.~once a large enough number of DM and $\eta$ particles have been created through the connector, the HS thermalizes and a dark thermal bath forms.
Following the arguments of Ref.~\cite{Chu:2011be,Chu:2013jja}, for example for $\alpha_X\sim 0.1$, one can check numerically that dark thermalization happens around $T\sim\mathcal{O}$(TeV)  if $\lambda_m$ is larger than $\sim 10^{-12}$. 
 Notice that the thermalization process increases the HS number density at the expense of reducing the average energy per dark particle.

\item[$(b)$] As mentioned above, the energy transfer stops before $T'$ reaches $\ma$ because at these temperatures connector processes are suppressed by the smallness of both the SM fermion Yukawa couplings and the mixing angle. This occurs at $T\sim{\cal O}(10)$\,GeV. After this time,
both sectors are completely decoupled and the corresponding entropies are conserved separately. 
In this epoch, since both sectors are relativistic, their temperatures are inversely proportional to the scale factor $a$, and therefore $T'\propto T\propto 1/a$. Due to the same reason, the yield $Y_A$, i.e.~the ratio of DM number density to SM entropy density, is basically constant. 

\item[$(c)$] Once $T'$ drops below $\ma$, DM becomes non-relativistic, the yield is no longer constant and, as long as the HS interaction is in thermal equilibrium, it becomes more and more Boltzmann suppressed. Using entropy conservation in the HS, it is possible to conclude that the exponential suppression factor leads to a milder dependence of the dark temperature on the scale factor than before, namely,  $T'$ scales as the inverse of the logarithm of the scale factor \cite{Carlson:1992fn}, $T'\propto 1/\log a$. As a result of this, while the visible sector cools down due to the Universe expansion as $1/a$, the HS cools down in a much slower way, i.e.~$T'/T$ increases as $\sim a/\log a$. This stems from the fact that chemical equilibrium still holds and consequently the 3-to-2 reaction `heats' the dark plasma. 

If the production process were to start from thermal equilibrium, i.e.~$T'/T=1$ at early times when $T'\gg \ma$, $T'/T$ would become much larger than one subsequently, in particular at freeze-out. If instead, as we assume here, $T'/T$ is much smaller than one at $T'\gg \ma$, even if it increases afterwards, it can remain smaller than one at freeze-out. In this way we avoid structure formation problems, i.e.~there is no need any more of establishing kinetic equilibrium between HS and SM bath, which as argued above  is otherwise in general difficult to satisfy.

\item[$(d)$] At last, the interaction rate of the 3-to-2 process becomes smaller than the Hubble expansion rate, leading to departure from chemical equilibrium. The total number of DM particles freezes and, as usual, the average momentum of DM simply redshifts with the Universe expansion. Moreover, as DM momentum is inversely proportional to the scale factor, its average kinetic energy goes like the square of it. Correspondingly, after the non-relativistic freeze-out, the DM temperature decreases faster than the visible one as $T' \propto T^2$. 
\end{enumerate}

\begin{figure}[t!]
\centering\includegraphics[width=0.46\textwidth]{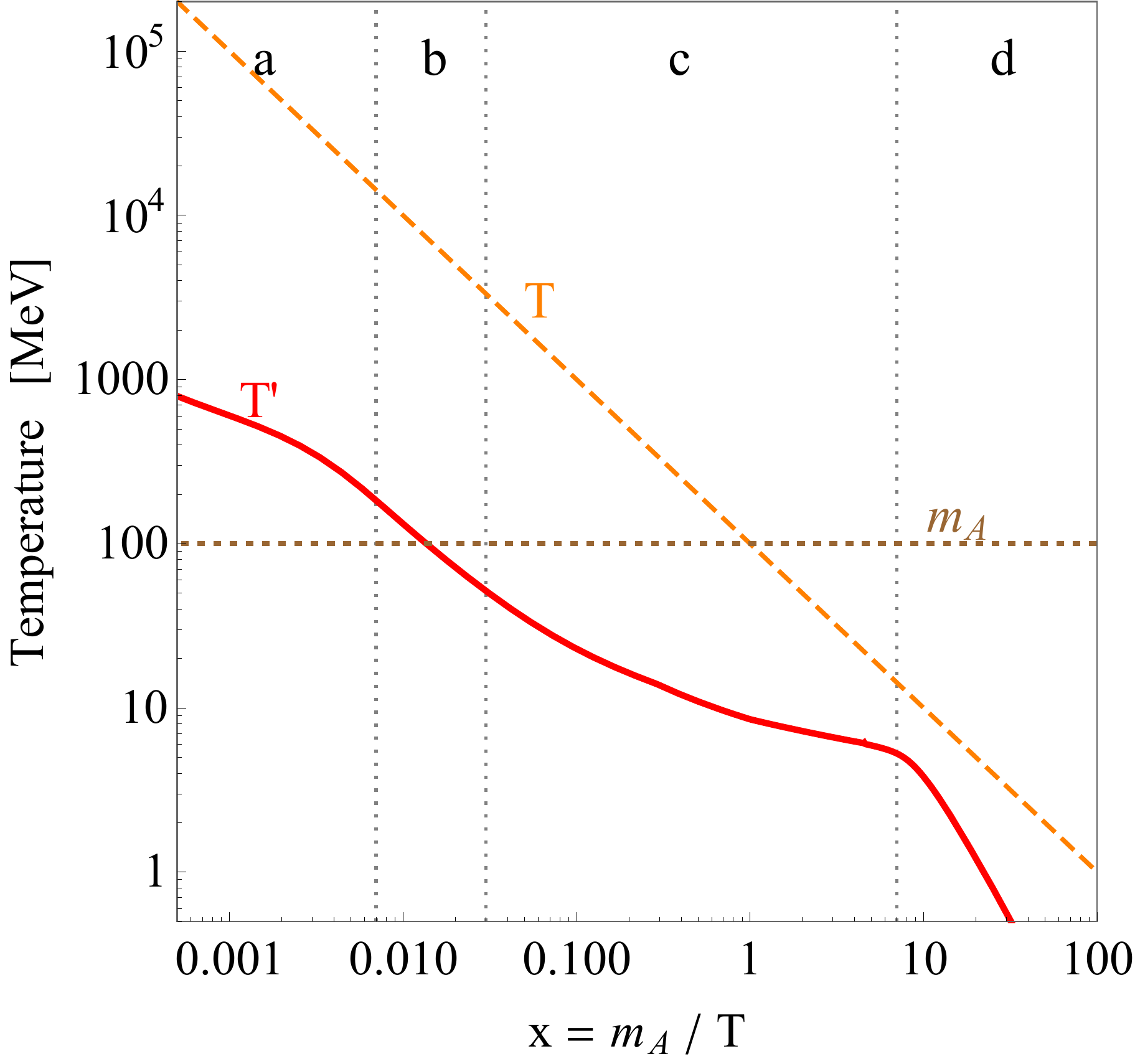}
\centering\includegraphics[width=0.50\textwidth]{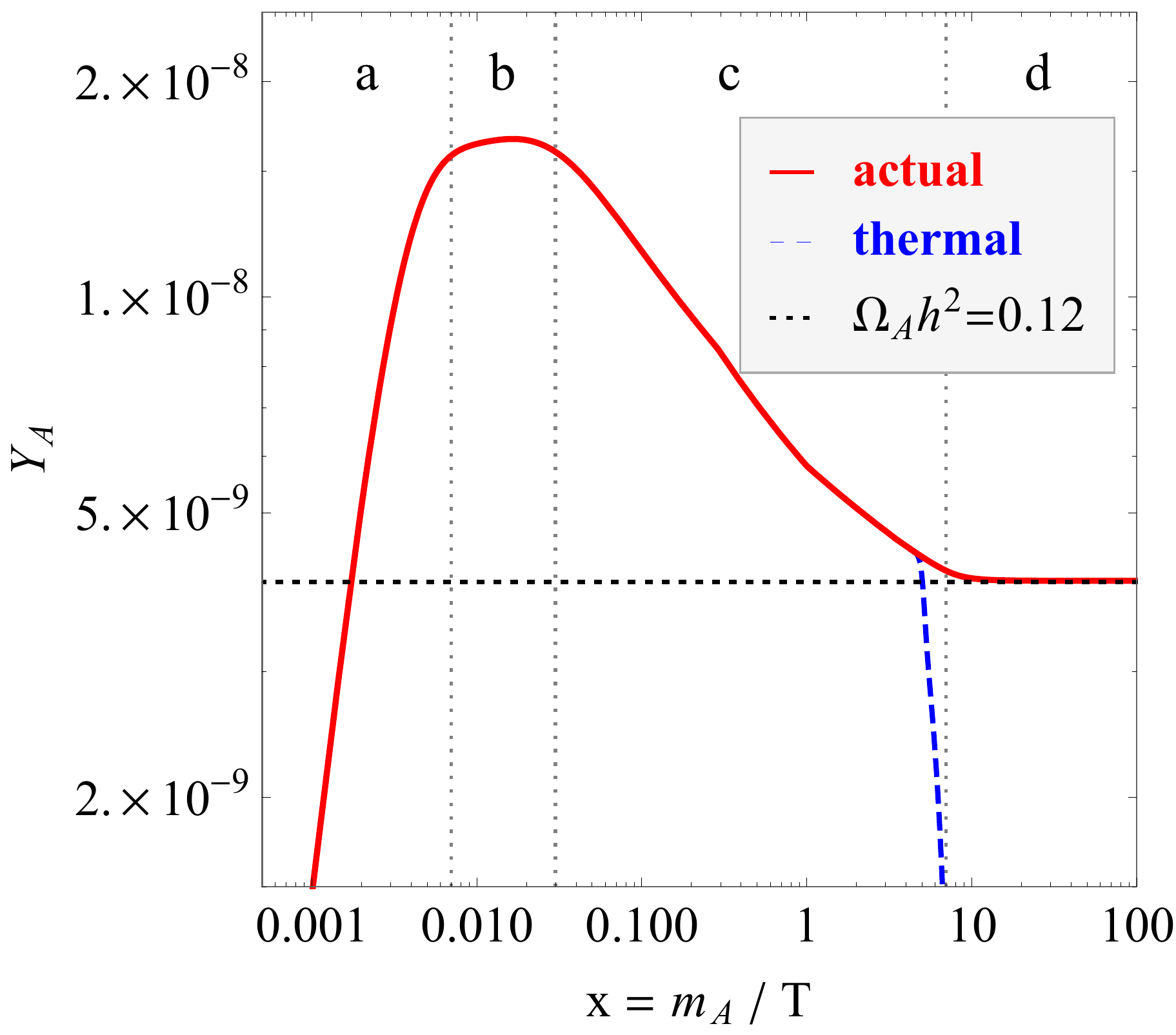}
\caption{Dark Freeze-out with $T'<T$ for the benchmark point $\ma = 100$\,MeV, $\meta = 300$\,MeV, $\alpha_X = 0.08$ and $\lambda_m= 4.3\times 10^{-12}$. Left panel: Evolution of the temperatures of the visible and hidden sectors, labeled as $T$ and $T'$, respectively. The horizontal dotted line corresponds to a temperature equal to the DM mass. Right panel: The evolution of the ratio  DM number density to SM entropy density yield, $Y_A = n_A/s$. The red solid curve corresponds to the \emph{actual} evolution, while the dashed blue line corresponds to the equilibrium value $n^\text{eq}(T')/s(T)$ (labeled as \emph{thermal}). We also show the yield corresponding to the observed DM abundance (dotted black line, labeled as $\Omega_A h^2$ = 0.12~\cite{Ade:2015xua}). }
\label{fig:caseB}
\end{figure}

The evolution of the DM yield is shown in Fig.~\ref{fig:caseB} for a particular benchmark point with $\ma=100$\,MeV that matches the DM relic density as measured by {\tt Planck}~\cite{Ade:2015xua}. There, we depict the evolution of dark temperature and DM yield on the left and right panels, respectively. $(a)$ On the first place, for $x \lesssim 7\cdot 10^{-3}$, i.e.~$T \gtrsim 14$\,GeV, the energy transfer from the visible sector to the HS takes place and as a result the dark temperature increases with respect to the visible one. Correspondingly, on the right panel we observe that the DM abundance grows. $(b)$ In between $x \sim 7\cdot 10^{-3}$ and the time when DM becomes non-relativistic, at $x \sim 3\cdot 10^{-2}$ (i.e.~$T'\sim \ma/3$), both the visible and the dark temperatures have a common slope signalizing that they are proportional to each other. During this period the $Y_A$ yield is basically constant. $(c)$ When the DM becomes non-relativistic, at about $T'\sim \ma/3$ and before the freeze-out, or in the plot for $3\cdot 10^{-2} \lesssim x \lesssim 7$, the DM abundance becomes Boltzmann-suppressed and $T'$ only decreases logarithmically due to entropy conservation. In fact, the temperature ratio $T'/T$ changes from 1/100 to about 1/3 during this epoch. $(d)$ Finally, after $x\sim 7$, the chemical decoupling takes place as the actual abundance starts to deviate from the equilibrium value and DM freezes-out. Likewise, $T'\propto T^2$ as argued before.

Beyond this benchmark example, to determine the allowed parameter space it is convenient to proceed as follows. As described above, after the energy transfer from the visible to the HS stops, typically at $T\sim {\cal O}(10)$\,GeV, the ratio of entropies  between visible sector and HS, $\xi \equiv s/s'$, becomes a constant only depending on the portal coupling.  Thus, it is convenient to take this entropy ratio as one of the input parameters. From it, one can calculate the temperature ratio between the two sectors when the DM is relativistic
\begin{equation}
\xi = \frac{ g_{*s}(T)}{10}\left(\frac{T}{T'}\right)^3,
\label{xiTTprime}
\end{equation}
 where $g_{*s}(T)$ is the relativistic degrees of freedom contributing to the visible sector entropy at a temperature $T$, while the factor $10$ corresponds to relativistic degrees of freedom in the HS (three for each spin-1 boson and one associated to the scalar $\eta$). Furthermore, using entropy conservation in the HS as well as the observed relic abundance, it is possible to prove that the DM temperature at the freeze-out~\cite{Carlson:1992fn} is
\begin{equation}
\label{3to2:decouple}
T'_\text{FO} \simeq (3.6 \text{\,eV})\cdot \Omega_A h^2\cdot \xi\,.
\end{equation}
For instance, in Fig.~\ref{fig:caseB} the chosen benchmark corresponds to $\xi\sim 10^7$ and $T'_\text{FO}\sim 5$\,MeV. 

\begin{figure}[t!]
\centering
\centering\includegraphics[width=0.65\textwidth]{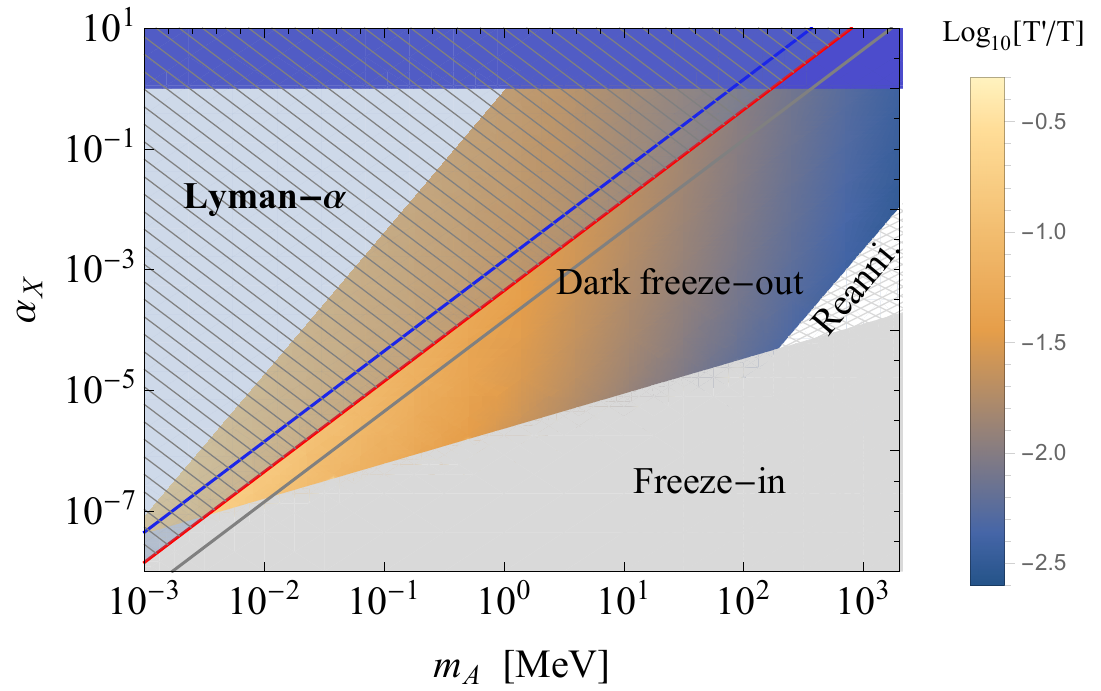}
\caption{Values of the temperature ratio $T'/T$  at $T'>m_\eta$ (and after the connector has ceased to create HS particles) required in order to generate the measured DM abundance via the 3-to-2 \emph{dark freeze-out} (colored band) under the assumption of no kinetic equilibrium between the hidden and visible sectors.
The gray, red and blue lines correspond to $\sigma_{AA}/\ma=0.1$, 1 and 10 cm$^2/$g, respectively, so that the hatched light-blue region is excluded by the cluster bound, $\sigma_{AA}/\ma>1$~cm$^2/$g. The left-top white region is excluded by Lyman-$\alpha$ data.  We also show the values of $m_A$ and $\alpha_X$ which lead to \emph{reannihilation} and \emph{freeze-in} regimes rather than dark freeze-out. See text for details. 
}
\label{ma-alpha-som2}
\end{figure}

That being the case, all relevant physical constraints on this regime can be expressed in the plane $\alpha_X-\ma$. Concretely, for a given gauge coupling $\alpha_X$ and a DM mass, one can determine the value of $\langle \sigma v^2\rangle_{3\to2}$ (since the dependence on $m_\eta$ is negligible, as explained above). From this value we can calculate the temperature $T_\text{FO}$ 
that is required to give the observed DM abundance, using Eq.~\eqref{FOconditionText} (see also Appendix~\ref{sec:BEandFO} for details). From $T_\text{FO}$, one can then determine $T'_\text{FO}$ with Eq.~\eqref{xpf},  and hence the value of $\xi$ needed from Eq.~\eqref{3to2:decouple}, and thus the temperature ratio when the HS particles are relativistic from Eq.~\eqref{xiTTprime}. Putting all these pieces together, the various constraints and the resulting allowed parameter space are shown in Fig.~\ref{ma-alpha-som2}.

As in Fig.~\ref{ma-alpha-som}, the gray, red and blue lines correspond to $\sigma_{AA}/\ma=0.1$, 1 and 10 cm$^2/$g, respectively. Similarly, the cluster exclusion and non-perturbative regions are given by the hatched light-blue region and blue top band, respectively. 
In addition, the upper white region turns out to be excluded by Lyman-$\alpha$ constraints, which we conservatively take as $T_\text{FO} \gtrsim 1$~keV~\cite{Viel:2013fqw}. Fig.~\ref{ma-alpha-som2} also shows that in order to satisfy the self-interaction  hypothesis and the relic density constraint one needs $\ma\lesssim 400$\,MeV. 

To sum up, the 3-to-2 mechanism, under the assumption of a HS cooler than the visible sector, arising naturally if the HS is created from the visible sector through a small portal,
solves the two problems found in the regime 3B. Concretely, there is no tension with cluster observations anymore and there are regions of the parameter space where the relic density is obtained with perturbative gauge couplings, as illustrated in Figs.~\ref{ma-alpha-som} and \ref{ma-alpha-som2}.

\vspace{0.1cm}

\subsection*{Regime 2: Reannihilation}
\label{subsec:reannih}
Fig.~\ref{ma-alpha-som2} shows the  fraction of the $m_A$-$\alpha_X$ plane which leads 
to the reannihilation regime. Clearly this regime is not compatible with large self-interactions as required by the small-scale problems. This happens because of the following reasons. As explained above, the (Yukawa suppressed) Higgs portal ceases to transfer energy from the visible sector to the HS at $T\sim \mathcal{O}(10)$\,GeV.
Thus reannihilation is possible only if $T_\text{FO}> 10$\,GeV. 
Not surprisingly, the  corresponding lower bound on $m_A$ for reannihilation to occur is not far below 10\,GeV (given the facts that in this regime $m_A \gtrsim T'_\text{FO} \gtrsim 10$\,GeV$\cdot(T'_\text{FO}/T_\text{FO})$ and that in order to have the correct relic density a large enough value of the ratio $T'_\text{FO}/T_\text{FO}$ is required).
For such value of $m_A$,  the self-interaction hypothesis can only be fulfilled with non-perturbative values of $\alpha_X$, 
whereas reannihilation occurs only for much smaller values of $\alpha_X$.
Notice that the line where the Lyman-$\alpha$ exclusion starts and the line where reannihilation takes place are parallel because they are both constraints on the same parameter, namely, $T_\text{FO}$. 

Nevertheless,  since the HS cross section required to yield the observed DM abundance is in general of the same order of magnitude as in the previous regime~\cite{Chu:2011be},
we must emphasize that in other models reannihilation could certainly work.
This requires that, unlike here, the process transferring energy from the visible sector to the HS are not suppressed at $T'\sim \ma \sim 100$\,MeV. This could be the case, for example, by considering a gauge portal, i.e.~a $Z'$, which is not Yukawa suppressed as the Higgs portal we have considered here. 

\subsection*{Regime 1: Freeze-in}
\label{subsec:FI}

The freeze-in regime is based on the assumption that the DM particle never thermalizes with any other particle.
Clearly, this can only happen if both the connector and the HS interactions are very small. 
Fig.~\ref{ma-alpha-som2} shows the region, labeled as `Freeze-in', where this assumption is fulfilled.
We see that either the values of $\alpha_X$ that are required to reproduce the observed relic density in this case are too small to have enough self-interactions,  or the mass must be very low, namely, $m_A$ of order few keV.
Actually, the Lyman-$\alpha$ constraint shown in Fig.~\ref{ma-alpha-som2} does not apply  here because there is no DM temperature in the freeze-in regime as dark thermalization does not happen. Nevertheless, the kinetic energy of DM particles -which simply redshifts after they are produced- is expected  to have a value close to the value of the visible temperature $T$. Thus the relevant constraint is simply  the usual hot DM one, $m_A\gtrsim 3$~keV~\cite{Viel:2013fqw}.
One finds therefore a small region of parameter space, with $m_{A}$ between 3~keV and 10~keV, where all constraints can be fulfilled. One can also check that this case is compatible with constraints on the amount of extra relativistic degrees of freedom. 

\subsection*{Regime 0: No Portal}
\label{subsec:noconnector}

Finally if the connector is so tiny that it has never had any practical impact, one can consider two possibilities, depending on whether the HS thermalizes or not.
If it does not thermalize, although there is a relic density we can not predict its value unless we address the physics responsible for the reheating process~\cite{Dev:2013yza,Kane:2015qea}. We will thus not consider this possibility any further. If it thermalizes, then the HS is characterized by a temperature $T'$ which could be smaller or larger than the visible sector one. If it is smaller, this leads to a viable scenario, which, as soon as the HS thermalizes, proceeds exactly in the same way as for the out-of-equilibrium dark freeze-out scenario above.


\section{Scenario with a Light Mediator}
\label{sec:LightMediator}

In this Section we consider the light mediator case, $\meta\ll \ma$, 
examining the various DM production regimes one-by-one as in the previous Section.
As discussed in Section~\ref{sec:SS}, in this case large self-scatterings can be achieved via non-perturbative effects due to the lightness of the mediator $\eta$. 

Three preliminary remarks are important to keep in mind all along this Section.  First,  as discussed in Section \ref{sec:SS}, in the light mediator scenario, having large enough self-interactions typically requires a fairly low mediator mass, below $\sim 100$\,MeV, and a larger DM mass from far below GeV to tens of TeV.
Second, as the mediator is lighter than the DM particle, 
DM can annihilate directly into a pair of mediator particles or to a $\eta$ and a DM particle, thus 2-to-2 processes always dominate over 3-to-2 annihilations. Third,  as we will see in detail below,
stringent constraints on the size of the Higgs portal, and thus on the strength of the connector processes, apply for low mediator  masses.

\begin{figure}[t]
\centering\includegraphics[width=0.47\textwidth]{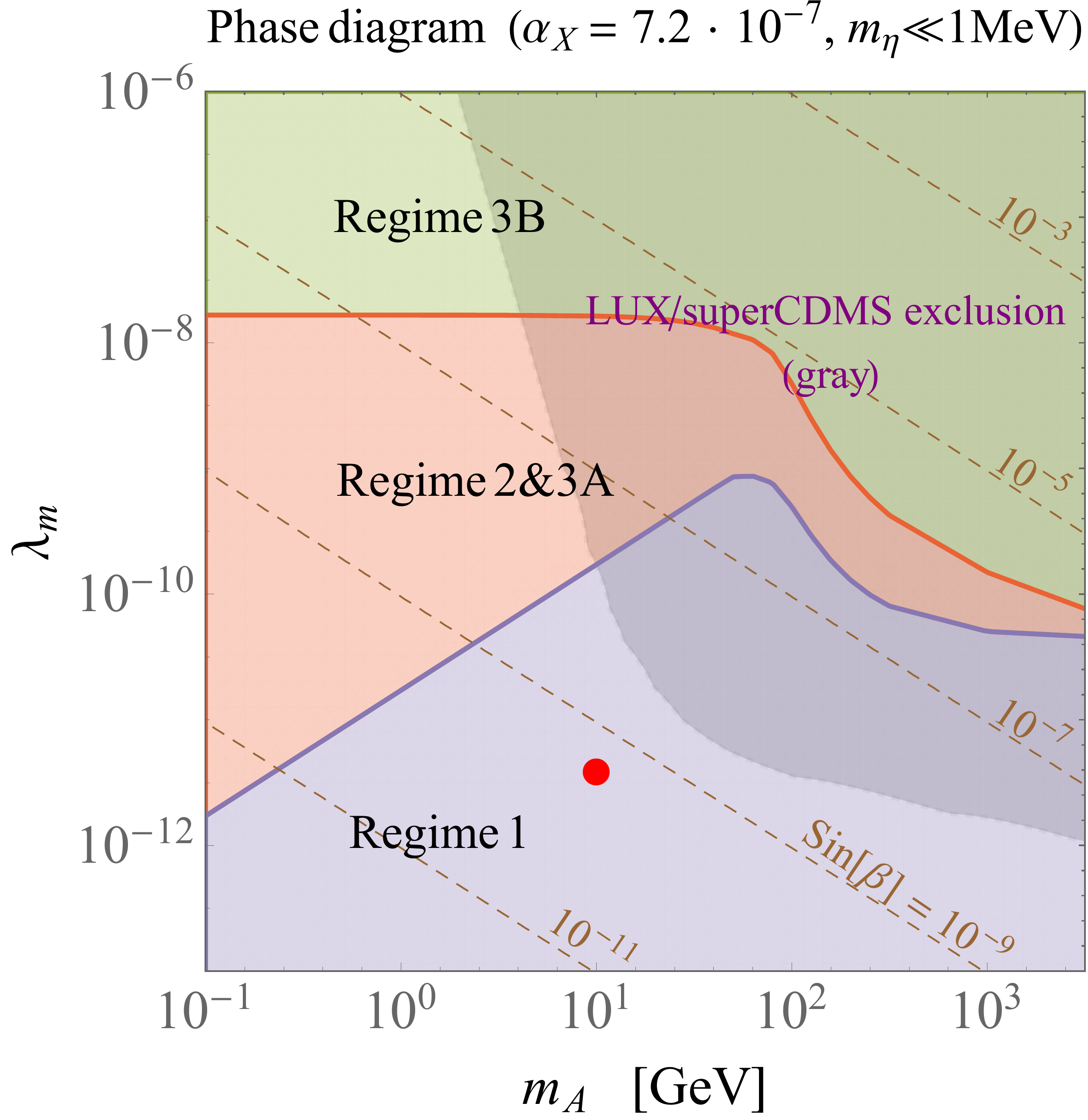}
\centering\includegraphics[width=0.47\textwidth]{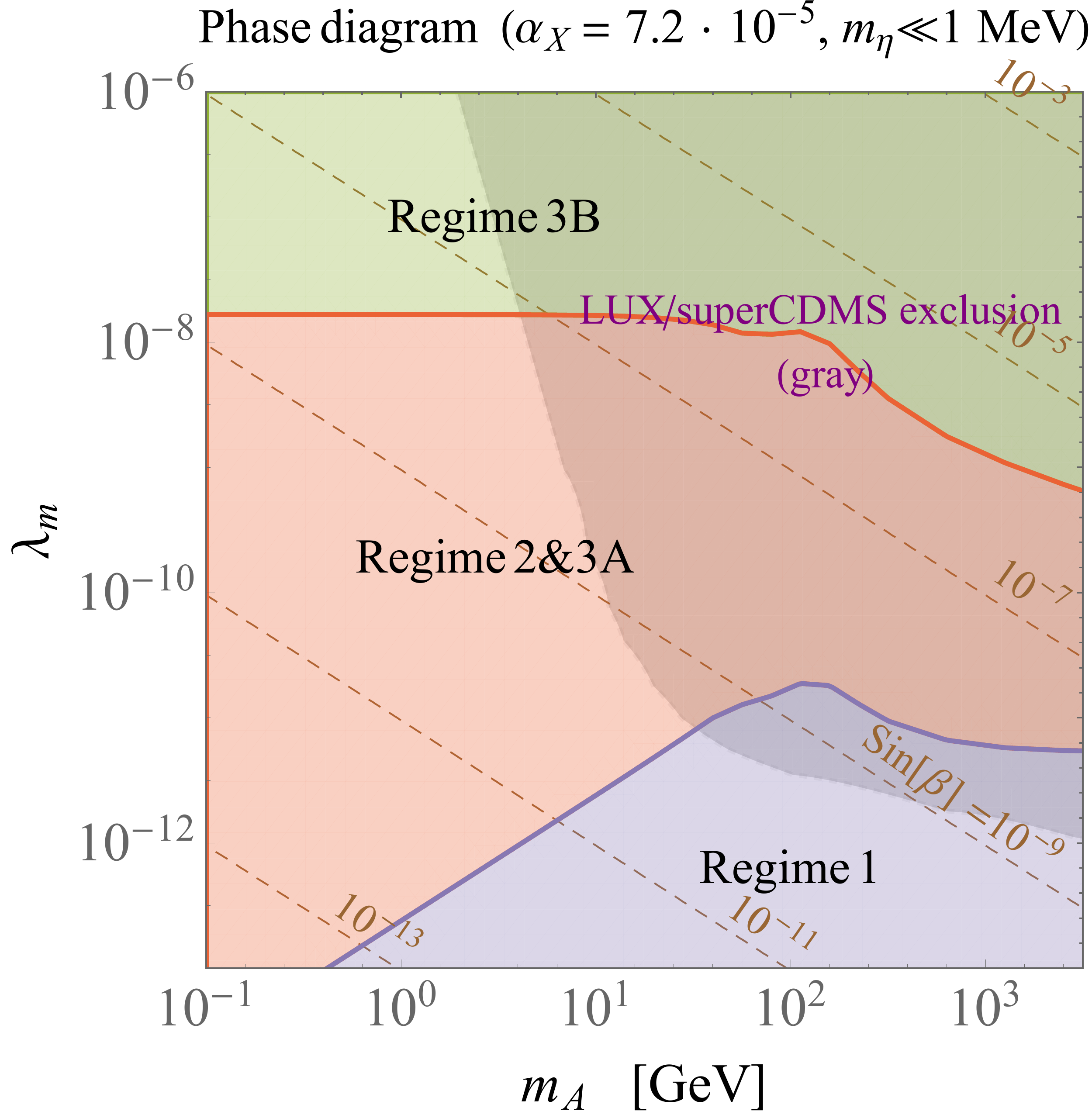}
\caption{Phase diagrams for two fixed values of $\alpha_X$, assuming $m_{\eta} \ll m_A$. The parameter areas in various colors correspond to various regimes defined in Section~\ref{sec:DMrd}. The gray area is excluded by DM direct detection bounds from {\tt LUX}/{\tt SuperCDMS}. The red point gives the parameter set used in  the discussion of the freeze-in regime in the text.}
\label{fig:LightMESA}
\end{figure}
  As mentioned in Section \ref{sec:DMrd}, for $m_\eta \ll m_A$ the relic density  depends only mildly on $m_\eta$ and hence the different regimes can be sketched in a phase diagram.  This was done in Fig.~\ref{phasediagram}, which shows, for $\ma=0.1$ and 10\,GeV, the production regimes that apply  as a function of the connector interaction $\lambda_m$ and HS interaction $\alpha_X$.
 Similarly in Fig.~\ref{fig:LightMESA}, we give for two fixed values of $\alpha_X$, the regime that applies as a function of the DM mass and connector interaction $\lambda_m$. In the figure, we also display the contour lines corresponding to constant values of $\sin\beta$, as given by $\sin\beta \simeq 2\lambda_m m_Av/ (g_X m_h^2)$. Note that for $m_A\ll m_h$, the connector processes are dominated by the decay of the SM scalar to HS particles, such as $h\rightarrow A_LA_L$ where the $L$ index refers to their longitudinal part, as one can expect from the equivalence theorem.
As a result, the connector processes can be characterized by $\lambda_m$ only (or equivalently, $g_X \sin\beta/m_A$), as shown by Figs.~\ref{phasediagram} and~\ref{fig:LightMESA}. For $m_A\gg m_h$ instead, the connector processes are dominated by pair annihilations of SM particles into HS particles.

\subsection*{Regime 4: Portal Interaction Freeze-out}

The usual freeze-out regime could a priori be realized in the HVDM model from DM annihilations to SM particles via the Higgs portal.\footnote{While this possibility has been studied for $m_\eta$ of electroweak mass scale in Ref.~\cite{Hambye:2008bq}, here we are concerned with the light mediator case, i.e. $m_\eta\lesssim {\cal O}(100)$\,MeV.} However,  with a light mediator, as for the 3-to-2 case above, this scenario is not viable in the context of self-interacting DM. 

If DM is light, say $m_A\lesssim 1$\,GeV, this regime does not work for the same reasons  as the ones stated in Section~\ref{sec:NoLightMediator}. That is, the annihilation cross section 
via Higgs portal is suppressed by the smallness of both the mixing angle $\sin\beta$ and the Yukawa couplings of light SM fermions. As a result, the DM annihilation cross section to the visible sector is always well below the canonical thermal value. Moreover, the self-interaction  hypothesis requires a large value of $\alpha_X$ which makes the annihilations channels into SM particles subdominant with respect to annihilations into HS particles.
 
In the opposite case, when $m_A\gtrsim 1$\,GeV, this regime is excluded by 
direct detection experiments. This stems from the fact that in this case the DM-nucleon elastic scattering cross section is dominated by the $t$-channel exchange of $\eta$ particles, which is greater by a factor of $(m_h/m_\eta)^4\,\gtrsim 10^{12}$ with respect to the Higgs exchange 
(for a more detailed discussion see Eq.~\eqref{DD} below).
Such an enhancement could be compensated by considering tiny values of $\lambda_m$ or $\alpha_X$. But this is not an option because  freeze-out via portal interaction requires fairly large values of the connector while self-interactions require rather large values of $\alpha_X$. 

\subsection*{Regime 3B: Dark Freeze-out with $T' = T$}
\label{case:VIA}

\begin{figure}[htbp]
\centering\includegraphics[width=0.5\textwidth]{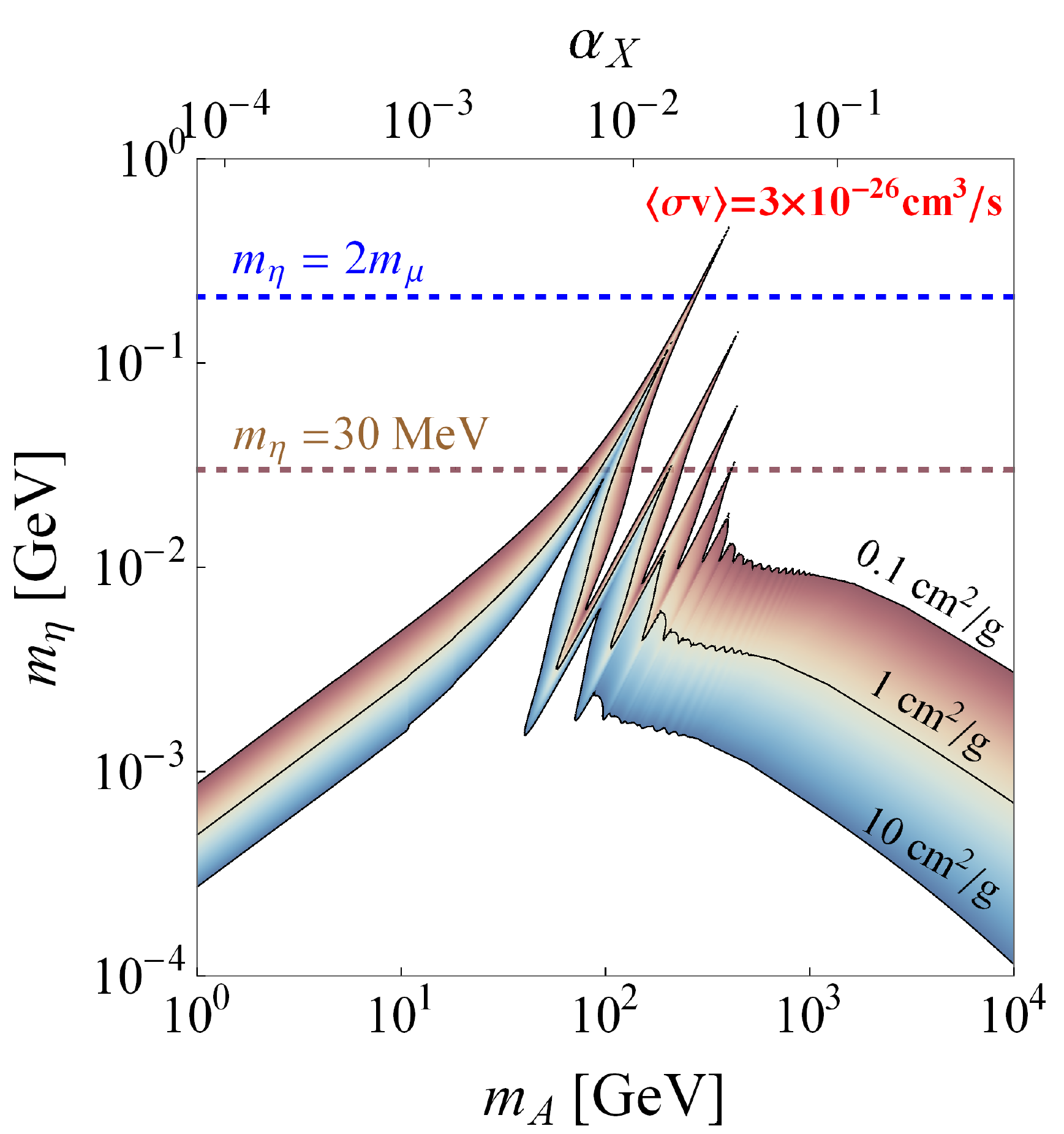}
\caption{Dark matter self-interactions for regime 3B.  Values of $\meta$ and $\ma$ which yield $\sigma_T/\ma$ in the range of $0.1-10$ cm$^2$/g (top-down), assuming a $v\sim 10$~km/s velocity.  The upper $x$-axis gives the value of $\alpha_X$ which leads to observed DM relic density, according to Eq.~(\ref{sigma22LM}). }
 \label{fig:caseC}
\end{figure}

In this regime DM annihilates to HS particles, through $AA\to \eta\eta$ or $\eta A$. The non-relativistic cross section of these annihilation channels is given by
\begin{equation}
\langle\sigma v\rangle\simeq \frac{11\,\pi}{72}\frac{\alpha_X^2}{\ma^2} + {\cal O}\left(\frac{\meta^2}{\ma^2}\right)\,.
\label{sigma22LM}
\end{equation}
Here we do not take into account the Sommerfeld enhancement induced by the lightness of $\eta$ because in this sort of scenarios it only leads to a small correction at freeze-out when $v\sim 1/3$, as proven in Ref.~\cite{ArkaniHamed:2008qn}.
Once the DM abundance is fixed to $\Omega_A h^2\sim 0.12$~\cite{Ade:2015xua}, the annihilating cross section of Eq.~\eqref{sigma22LM} implies a one-to-one correspondence between $\alpha_X$ and $\ma$.
Trading $\alpha_X$ for $m_A$ through this correspondence, 
the self-scattering $\sigma_T/ \ma$ cross section can be expressed as a function of only two parameters, $\ma$ and $\meta$. This is shown in Fig.~\ref{fig:caseC}, assuming
as in Section~\ref{sec:SS} above,  
 the value of the DM velocity which typically applies to dwarf galaxies, $v=10$~km/s.
For $m_A$ within the GeV to few TeV range, in order to have enough self-scattering, the mediator mass $\meta$ must be in the range $\sim10^{-3}-10^{-2}$\,GeV, unless one lies on a resonance. In that case, values of $m_\eta$ up to $\sim 300$\,MeV are possible.

Note that the  cluster bound discussed in Section \ref{sec:SS} (i.e.~$\sigma_T/m_A \lesssim$ 1\,cm$^2$/g for typical cluster velocity $v \sim 1000$~km/s) are relevant only for 
$m_\eta$ below a few keV, thus they do not appear in Fig.~\ref{fig:caseC}.

Given the fact that in this regime the HS is assumed to thermalize with the visible sector, the connector interaction must be sizable, as shown in Figs.~\ref{phasediagram} and~\ref{fig:LightMESA}. Consequently, relevant experimental constraints apply. Besides the constraint coming from the invisible decay width of the Higgs boson that
may apply in this case  (see Appendix~\ref{sec:Higgsdec}) as well as the one from  
the electron beam-dump experiment E137 at SLAC which requires $\sin\beta \lesssim 10^{-2}$ ~\cite{Bjorken:2009mm,Batell:2014mga}, one can distinguish: 

\begin{enumerate}[1.]
\item 
\textbf{DM direct detection:} 
The direct detection constraints, as given in particular by 
the {\tt LUX}~\cite{Akerib:2013tjd} and the {\tt SuperCDMS}~\cite{Agnese:2014aze} collaborations, turn out to be very stringent whenever $m_A$ is larger than a few GeV, due to the $t$-channel light mediator enhancement of the corresponding cross section. More precisely, in our model, DM-nucleon elastic scattering is spin-independent and mediated by the $\eta$ and Higgs particles. In the limit of $m_h\gg \meta$, the corresponding cross section is
\begin{equation}
\sigma_{AN} \simeq \frac12\frac{\alpha_X\, f_N^2\, s^2_\beta\, c^2_\beta\, \mu_{AN}^2}{\left(\meta^2+2m_NE_R\right)\left(\meta^2+4\mu^2v^2\right)}\,,
\label{DD}
\end{equation}
where $m_N$ and $\mu_{AN}$ are the nucleon mass and the DM-nucleon reduced mass, respectively. Here, $E_R\sim{\cal O}$(10)~keV is the typical nucleon recoil energy, and $f_N\sim0.3\,m_N/v_H$ is the effective Higgs coupling to the nucleons. Thus for low $m_\eta$ the direct detection cross section are largely enhanced.

\item 
\textbf{Big Bang Nucleosynthesis:} 
Since along the regime considered here the HS thermalizes, the $\eta$ particles are present in large amounts. 
As a consequence, the late decays of $\eta$ could spoil the prediction of BBN abundances if $m_\eta$ is large enough to have a decay product with sufficient energy, roughly $m_\eta\gtrsim 1$\,MeV. In that case, the $\eta$ scalar dominantly decays into a pair of SM leptons, which can not produce hadrons given the $\eta$ masses under consideration (see  Fig.~\ref{fig:caseC}).
Accordingly, BBN observations require the $\eta$ particle to have either a lifetime shorter than $\sim 10^4$\,s or to have a tiny relic abundance before it decays, $\Omega_\eta h^2\lesssim 10^{-5}$~\cite{Jedamzik:2009uy}.

Note that for $m_\eta \gtrsim 2\,m_\mu$ (which is only possible along the fine resonances shown in Fig.~\ref{fig:caseC}), the BBN bound is less stringent, since the decay channel to muon-antimuon pair is faster.

\item \textbf{DM annihilations at CMB and low redshifts:} 
It is well known that light thermal DM annihilating via a $s$-wave into SM particles is strongly constrained by both CMB measurements and indirect DM searches at low redshifts~\cite{Steigman:2015hda,Slatyer:2015jla}. In particular, since $\eta$ is unstable for $m_\eta\gtrsim 1$\,MeV, 
DM annihilations produce $\eta$ particles which subsequently decay into electrons. Consequently, thermal freeze-out scenario 
 has been excluded for $\ma \lesssim 12$\,GeV by {\tt Planck} data~\cite{Madhavacheril:2013cna}.
 {\tt Fermi-LAT} data gives similar constraints.
For $m_\eta<2\,m_e$  these particular CMB constraints might not apply because the $\eta$ particle can have a lifetime much longer than the time where the last scattering occurred~($\sim 10^{12}$\,s).

\item \textbf{CMB bounds on energy injection from decaying particles:} If the lifetime of $\eta$ (i.e. $\tau_\eta$) is larger than $10^{12}$\,s, it is possible to constrain the abundance of $\eta$ particles at the time of recombination because its decay products might distort the CMB spectrum via energy injection in Cosmic Dark Ages. In this work, we take the corresponding constraints from Ref.~\cite{Slatyer:2012yq}. For instance for $\tau_\eta \sim 10^{17}$\,s, which is about the age of the Universe, the abundance $\Omega_\eta$ must be smaller than $10^{-8}$ after recombination.   

\item \textbf{$\boldsymbol{X}$-ray emission bound:}
Whenever the lifetime of the $\eta$ particle is larger than the age of the Universe, which may occur for $m_\eta<2\,m_e$, one has to make sure that the leading decay channel $\eta \rightarrow \gamma\gamma$ does not lead to  a too large cosmic $X$-ray excess.
The corresponding decay width is 
\begin{equation}
\Gamma(\eta\to\gamma\gamma) = \sin^2\beta\cdot\Gamma(h\to\gamma\gamma)\big|_{\mh\to \meta}\,,
\label{etaWidth}
\end{equation}
where $\Gamma(h\to\gamma\gamma)$ is the Higgs diphoton decay width (see e.g. Ref.~\cite{Djouadi:2005gi}). 
A conservative bound can be taken as ~\cite{Essig:2013goa,Boddy:2015efa,Riemer-Sorensen:2015kqa}
\begin{equation}
\tau_\eta \gtrsim 10^{28}~{\rm sec}\times \df{\Omega_\eta h^2}{0.12}\,.
\label{diffuse}
\end{equation}
Then, Eq.~\eqref{etaWidth} leads to
\begin{equation}
\label{caseD:Xray}
\left( \frac{\sin\beta}{10^{-9}} \right)^2\cdot \left(\frac{\meta}{\text{1~keV}}\right)^3\cdot \frac{\Omega_\eta h^2}{10^{-6}} \lesssim 1\,. 
\end{equation}

\end{enumerate}

From the five constraints above and the thermalization condition ($T=T'$) which has to hold for this regime, one of the following pictures emerges depending on whether $\eta$ decays into $e^+e^-$ or $\gamma\gamma$.

\begin{figure}[t]
\centering\includegraphics[width=0.49\textwidth]{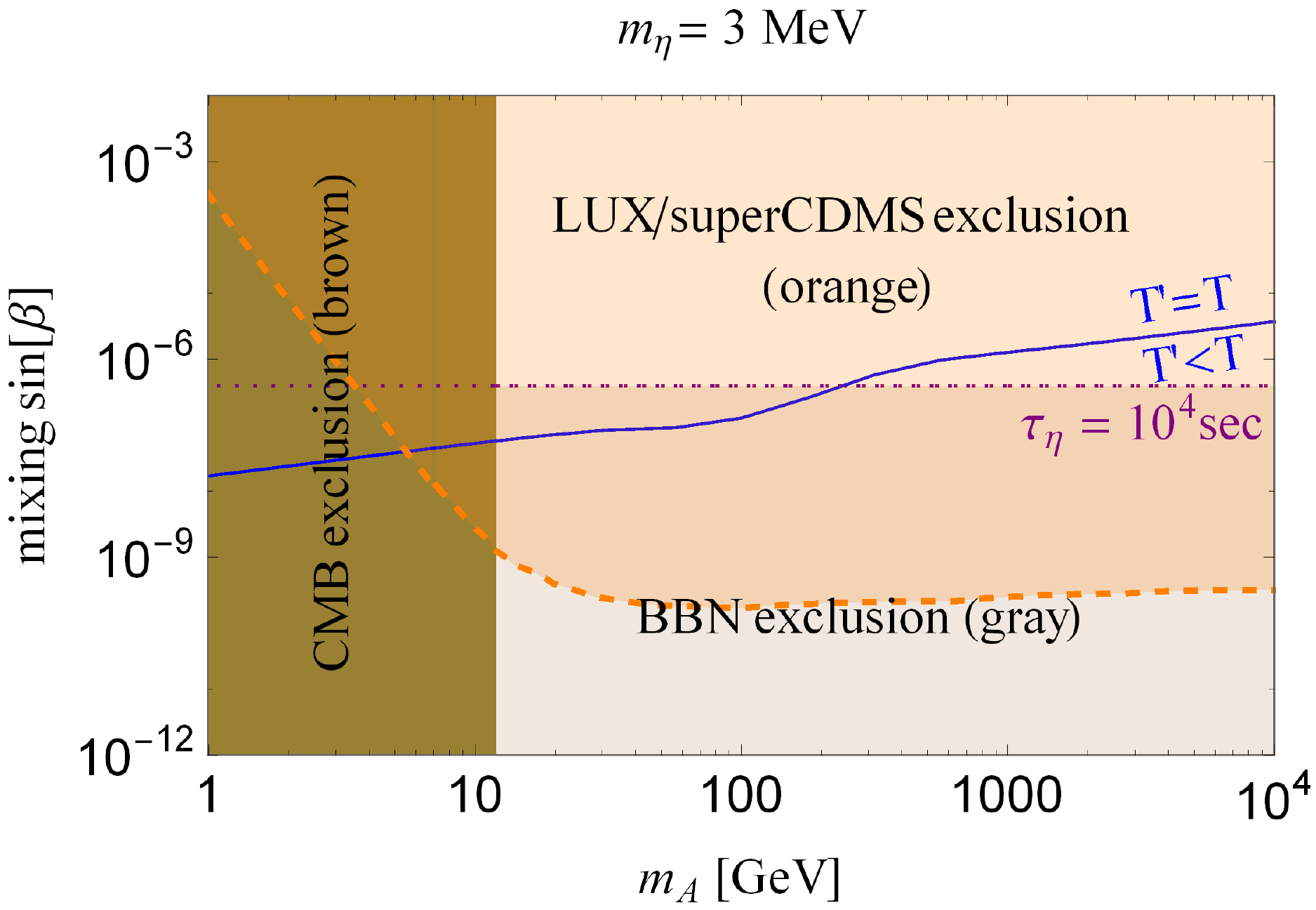}
\centering\includegraphics[width=0.49\textwidth]{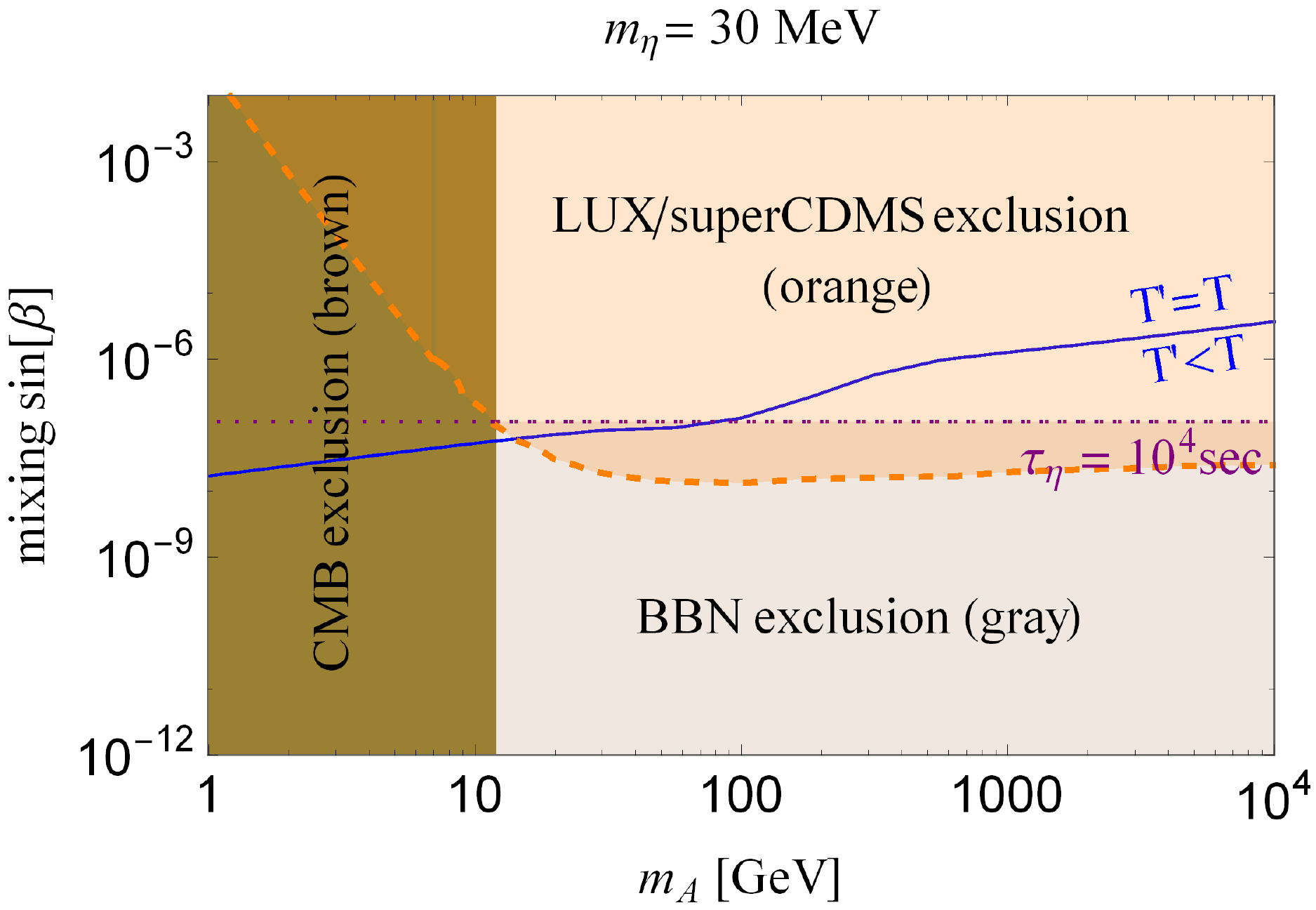}
\caption{ Constraints which hold on $m_A$ and $\sin\beta$ in the $T'=T$ dark freeze-out regime, for $m_\eta= 3$\,MeV~(left panel) and $30$\,MeV~(right panel). We show the direct detection upper bound on $\sin\beta$, the BBN lower bound on $\sin\beta$ and the constraint of thermalization of the HS with the visible sector, as labeled in the figure. Also, the left brown band corresponds to the lower limit on the DM mass coming from CMB constraints on annihilating DM.}
 \label{fig:caseCp}
\end{figure}

\begin{itemize}
\item 
We consider first the case when $\eta$ decays mainly into electron-positron pairs, therefore we take $2\,m_e<m_\eta<2\,m_\mu$. 

As explained above, BBN requires the $\eta$ particle to have either a lifetime shorter than $\sim 10^4$\,s or to have a tiny relic abundance before it decays, $\Omega_\eta h^2\lesssim 10^{-5}$~\cite{Jedamzik:2009uy}. It can be checked that the latter option is not possible because  it is in contradiction with the relation $ Y_\eta /Y_A \gg 1$, which naturally holds in this regime before the $\eta$ particles decay. Such relation stems from the fact that, 
when the DM particles become non-relativistic, their number density gets Boltzmann-suppressed 
 in contrast to the $\eta$ number density.
The BBN constraint is
$\tau_\eta\lesssim 10^4$\,s, which translates into $\sin\beta\gtrsim 10^{-7}\sqrt{0.1\,{\rm GeV}/\meta}$. This bound along with other constraints are shown in Fig.~\ref{fig:caseCp}, which summarizes what happens in this regime when $m_\eta=3$ and 30\,MeV. First, in this figure we show the DM direct detection upper bound on the mixing angle. As is clear from the plots, decreasing the mass of $\eta$ makes the bound more stringent. Second, we also depict the constraint from BBN ($\tau_\eta<10^4$\,s) which translates into a lower bound on the mixing angle.
Likewise, this lower bound becomes more severe for lighter $\eta$. Third, we also draw the line corresponding to the thermalization of both sectors, which gives another lower bound on the mixing angle. This line is barely affected by the mass $\meta$. Finally, since $\eta$ is short-lived,  the above lower bound on the DM mass of around 12\,GeV, coming from the CMB and low redshift  data, also applies here. 

The right panel of Fig.~\ref{fig:caseCp} shows that all these constraints exclude this regime for $m_\eta = 30$\,MeV. This conclusion is also true for lower masses of $\eta$ because decreasing $m_\eta$ makes the constraints more stringent, as shown in the left panel. Nevertheless, regions with larger $m_\eta$ are not ruled out  by the above constraints. However they correspond to the fine resonances shown in Fig.~\ref{fig:caseC}, which would lead to extremely large annihilation signals in DM indirect searches, such as {\tt Fermi-LAT}~\cite{Massari:2015xea} and {\tt Planck}~\cite{Ade:2015xua}, and therefore we do not consider that possibility in the sequel. 
  
\item Next we consider the cases in which the $\eta$ scalar decays into photons. Notice that $\eta$ can not be arbitrarily light because this would induce a too large self-interaction among DM particles. We therefore consider $100$~eV $<m_\eta < 2\,m_e$.
\begin{itemize}
\item[$(i)$] $m_A\,\gtrsim~10$\,GeV: Here  the DM-nucleon elastic scattering rate is significantly enhanced by the lightness of $\eta$ and consequently exclude this parameter region.
\item[$(ii)$] $m_A\lesssim10$\,GeV: In this case, we must consider three possibilities according to the lifetime of $\eta$. For $\tau_\eta < 10^{12}$\,s, DM annihilations produce $\eta$ pairs, which subsequently decay into photons. Because the annihilation cross section is expected to be of the order of few $10^{-26}$cm$^{3}/$s, this parameter region is excluded  from CMB bounds on DM annihilation, as explained above.  For $ 10^{17}$\,s $>\tau_\eta > 10^{12}$\,s, CMB bounds from energy injection during Cosmic Dark Ages apply  to the decay of $\eta$ and exclude this possibility because $\Omega_\eta\gg 1$ at recombination (given the lifetime under consideration). Finally, the possibility of $ \tau_\eta> 10^{17}$\,s is excluded by $X$-rays observations due to the fact that $\Omega_\eta\gg 1$, as explained above. 
\end{itemize}
\end{itemize}

To sum up, bounds from DM searches together with cosmological measurements basically exclude the large mixing ($T=T'$) scenario, as summarized in Fig.~\ref{fig:caseCp}.
Analogous conclusion have been obtained in various ways in other models.\footnote{For instance, in order to avoid these bounds, Ref.~\cite{Kouvaris:2014uoa} proposes a fermionic DM model, with the light mediator decaying into sterile neutrinos.}

\subsection*{Regime 3A: Dark Freeze-out with $T' < T$}
\label{sec:DarkFO}

Even if the problems encountered above with a light mediator are 
not precisely the same as the ones found
in Section~\ref{sec:NoLightMediator}, again a possible solution is to abandon the requirement of thermalization. This can be achieved by considering a smaller mixing between both sectors. Under this circumstance, the first possibility that one can consider is that the HS thermalizes with itself and undergoes a dark freeze-out, with $T' < T$. In practice, this  still leads to $Y_\eta \gg Y_A$ and to an annihilation cross sections smaller than few $10^{-26}$~cm$^{3}/$s (but still relatively large in order that the HS thermalizes with itself). 

Using this and after imposing the  five constraints listed in the previous subsection, we find that this scenario is arguably excluded except for  narrow parameter regions as shown in detail in Appendix~\ref{sec:AppendixC}. Moreover, the viable region left can be constrained and  could even be excluded by astrophysical searches of axion-like particles.  As this concerns only small corners of the parameter space we will not analyze further these possibilities.

\subsection*{Regime 2: Reannihilation}
\label{sec:Reanni}

From the phenomenological point of view, the reannihilation regime is analogous to the regime 3A, in the sense 
that it also requires a relatively large DM annihilation cross section  and also gives $Y_\eta \gg Y_A$.
As a consequence, we reach the same conclusion: this scenario is excluded, with the exception of a possible very narrow region, as argued in Appendix~\ref{sec:AppendixC}. 

\subsection*{Regime 1: Freeze-in}
\label{sec:freezein}

The freeze-in regime requires values of $\alpha_X$ and $\sin \beta$  small enough for the DM particle not to reach kinetic equilibrium. In this case, only the connector is responsible for the DM relic density and the $\alpha_X$ interaction does not play any role in it (i.e. the relic density constraint does not give any lower bound on the value of $\alpha_X$).  As a result of these facts, the direct detection constraints are easily fulfilled.
It will be shown below that at the same time this regime is compatible with a large enough value of $\alpha_X$ to satisfy the self-interactions hypothesis, provided the mediator is taken light enough to enhance sufficiently these self-interactions.
Moreover,  unlike in previous regimes, $\Omega_\eta h^2 \ll \Omega_A h^2$ naturally arises in this scenario, which makes the vector bosons $A$ the dominant DM component, and allows to satisfy easily  the other constraints which apply.

In the freeze-in regime, HS particles are created from the visible sector proportionally to the efficiency of the portal interaction. The dominant production processes have been shown already in Fig.~\ref{SMtoDM:channels}. For the present model, the freeze-in production of HS particles is infrared-dominated until this production stops. Thus a simple dimensional analysis implies that after the HS has frozen-in $Y_A/Y_\eta \sim T_{\text{dec},\eta}/T_{\text{dec},A}$, where $T_{\text{dec},i}$ stands for the temperature at which the production of the particle `$i$' becomes suppressed.\footnote{ For a detailed study of the DM freeze-in production mechanism, see Ref.~\cite{Blennow:2013jba}.}
For the light mediator $T_{\text{dec},\eta} \sim 10$\,GeV, as a result of the fact that at this temperature the Higgs portal interaction that produces $\eta$ scalars becomes Yukawa suppressed. On the other hand,
for $A$ the suppression of the production rate takes place at $T_{\text{dec},A} \sim \hbox{Max}[m_A$, 10\,GeV$]$.
As a result, at the end of the freeze-in, for DM lighter than ${\cal O}(10)$\,GeV
one gets $Y_\eta/Y_A \sim 1$, which implies $\Omega_\eta / \Omega_A \sim \meta/\ma$; whereas for 
heavy DM ($m_A\gtrsim {\cal O}(10)$\,GeV), one gets $Y_\eta/Y_A \sim m_A/(10~\text{GeV})$, or equivalently, $\Omega_\eta / \Omega_A \sim \meta/(10$\,GeV),  i.e.~this ratio simply scales linearly in $m_\eta$.
Hence, unlike in the previous regimes, freeze-in  naturally leads to $\Omega_\eta/\Omega_A\ll1$ provided that $m_\eta \ll m_A$.

From the relations for $\Omega_\eta/\Omega_A$ described above and the observed DM relic abundance, one can conclude that if $m_\eta > 2m_e$, then $\Omega_\eta h^2 > 10^{-5}$. Thus in order to avoid BBN constraints, $\tau_\eta$ should be less than $10^4$\,s in this case. However in freeze-in regime, the $\eta$'s are much more long-lived due to the smallness of the couplings. This forces the mediator mass to obey $m_\eta \lesssim 2m_e$.
We therefore focus our analysis on the $m_\eta \lesssim 2\,m_e$ case, where the CMB and $X$-ray constraints are relevant.  These constraints can in fact be satisfied easily here, if the $\eta$ abundance is very small after recombination. As just mentioned above, this can be  realized if $m_\eta \ll m_A$. In practice this requires $m_\eta \ll 1$\,MeV. 

\begin{figure}[t]
\centering\includegraphics[width=0.45\textwidth]{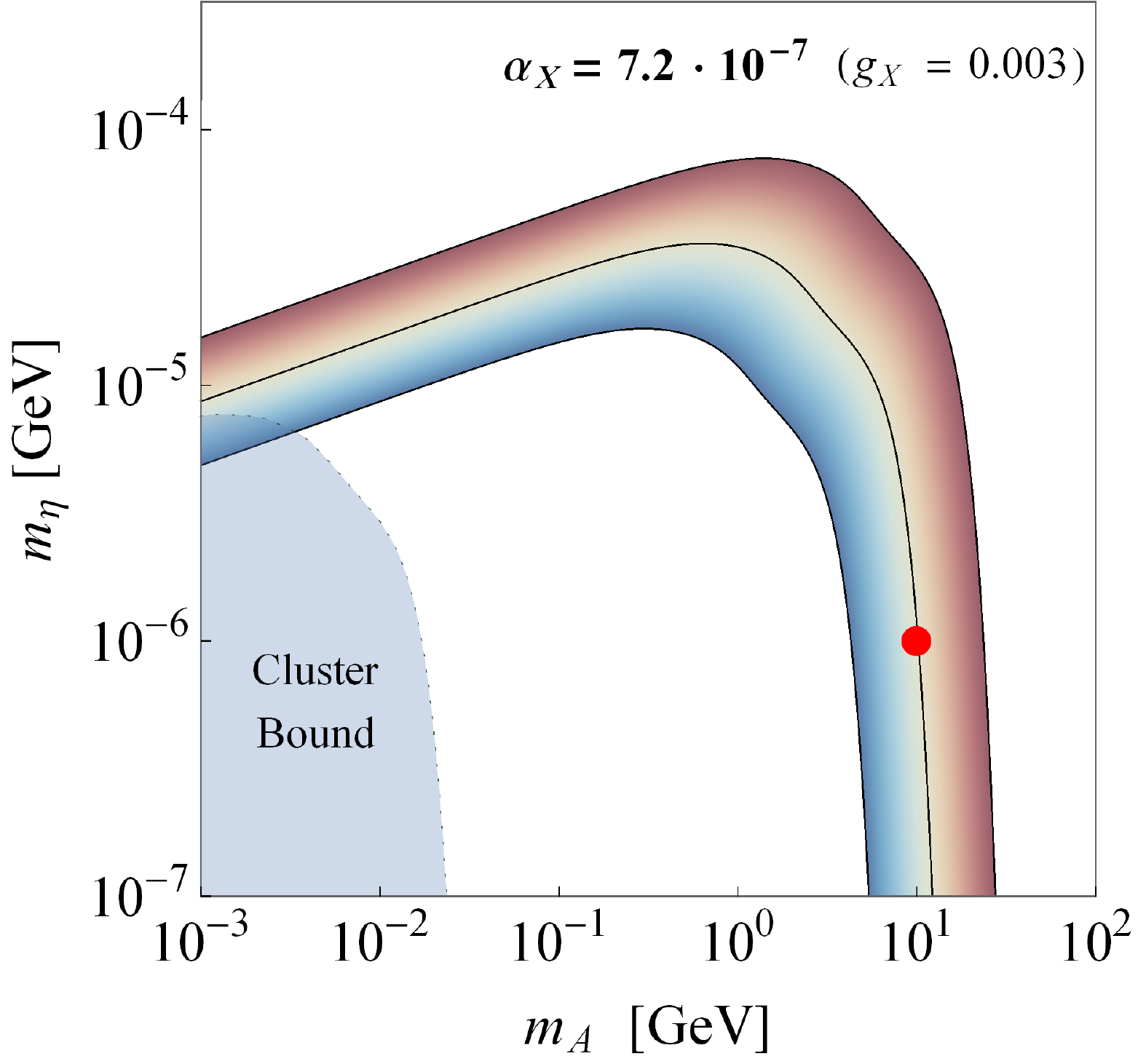}
\centering\includegraphics[width=0.45\textwidth]{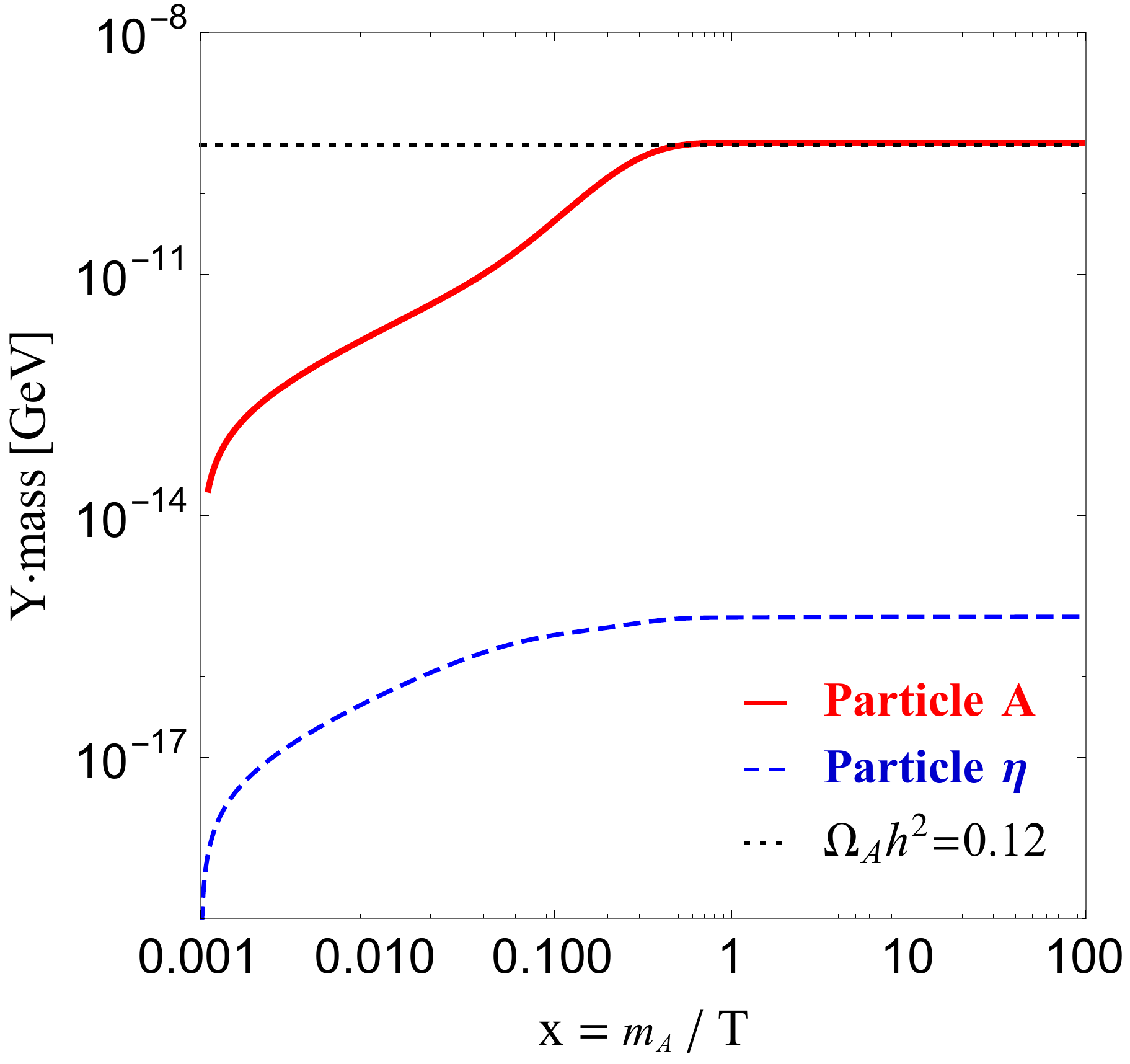}
\caption{Freeze-in regime. Left panel:  fraction of the $\meta-\ma$ plane which yields a value of $\sigma_T/\ma$ within the 0.1-10 cm$^2$/g range (top-down) for $v\sim 10$~km/s. 
The change of behavior around 
$\mathcal{O}(1)$\,GeV is due to the fact that when $m_A$ becomes larger than
$m_\eta/\alpha_X$, non-perturbative effects start to
dominate and the self-scattering cross section needs to be described
by classical regime results; whereas for smaller masses the Born approximation holds.
Right panel:  Evolution of the $A$ and $\eta$ particles abundance, given by $Y_{A,\eta} \cdot m_{A,\eta}$, as a function of the inverse of the temperature $T$, for a set of parameters corresponding to the red point in the left panel, i.e.~$\ma= 10$\,GeV, $\sin\beta= 3.9\cdot 10^{-10}$ ($\lambda_m = 3.8\cdot 10^{-12}$) and $\meta = 1$~keV. The difference between the $A$ and $\eta$ lines is mainly due to the large mass ratio $\ma/\meta$.
}
 \label{fig:caseD}
\end{figure}

In Fig.~\ref{fig:caseD} we show how this can be achieved for a particular set of parameters, which
gives  both a DM self-interaction in the range needed for solving the small-scale problems (left panel), and the correct relic abundance (right panel). In the left panel we have taken $g_X =3\cdot 10^{-3}$, whereas in the right panel we additionally assume $\ma= 10$\,GeV, $\sin\beta= 3.9\cdot 10^{-10}$ (i.e. $\lambda_m = 3.8\cdot 10^{-12}$) and $\meta = 1$~keV, where the latter value was chosen in order to satisfy the $X$-ray constraint of Eq.~\eqref{caseD:Xray}. The right panel of Fig.~\ref{fig:caseD} shows numerically that, as expected, the energy transfer becomes most efficient around the electroweak scale, and gradually stops at $T\sim$10\,GeV. The ratio of the two relic abundances one gets numerically in this case, $\Omega_A /\Omega_\eta$, is of the order ${\cal O}(10^6)$, which agrees well with our estimates above. 
The gray region in left panel of Fig.~\ref{fig:caseD}, denoted as `cluster bound', gives the DM self-interaction bound derived from galaxy cluster data, which is taken as $\sigma_T/m_A \lesssim$ 1\,cm$^2$/g with $v\sim 1000$~km/s.

%

\vspace{.1cm}

Regarding the $X$-ray constraint, notice also that since in the freeze-in regime the $\eta$ particle might be part of the DM relic density today (at a subdominant level), its decay could be detected by future $X$-ray observations.  Moreover, it can be shown that it might explain the 3.5~keV $X$-ray excess~\cite{Bulbul:2014sua,Boyarsky:2014jta}.

  Concerning direct detection, it is known that the $m_A\lesssim1$~GeV region is still below current experimental reach\,(see Fig.~\ref{fig:LightMESA}) but future experiments (e.g. {\tt Xenon1T}~\cite{Davide:2015bca} or {\tt LZ}~\cite{Akerib:2015cja}) will be able to probe this region. For sub-GeV DM masses, there would be in principle constraints from electron-recoil analysis~\cite{Essig:2011nj,Essig:2012yx,Hochberg:2015pha,Chen:2015pha,Essig:2015cda}, but here they are irrelevant anyway, because both $A$ and $\eta$ particles are only feebly coupled to electrons.

\subsection*{Regime 0: No Portal}
\label{sec:noportal}

In this regime, as there is almost no mixing between both sectors (or none at all), the $\eta$ scalar is
always stable and contributes to the observed DM relic density.
By taking it sufficiently light, one can easily make its relic density subdominant with respect to the vector gauge bosons $A$ and fulfill the self-interaction hypothesis. In addition, depending on whether the HS is thermalized or not, it either
goes through a dark freeze-out or just cools down with the Universe expansion, respectively. In both cases, one then only needs to make sure that the average kinetic energy per HS particle is small enough with respect to the visible sector temperature so that cosmological constraints, which are the only ones that apply here, are satisfied. 
 

\section{Conclusions}
\label{conclusions}

In summary, we have shown in a systematic way how the various possible DM production regimes,  where the DM sector is not in thermal equilibrium with the SM sector, can be relevant for solving the small scale DM problems.
In order to have large self-interactions (as required by small scale problems) and to accommodate the
relic density constraint, 
we performed this analysis invoking either a 3-to-2 annihilation to fix the DM abundance, or a light mediator
to enhance the self-interaction cross section through long-range quantum mechanical effects. 
Other constraints that are fully relevant for such scenarios come from BBN, CMB, $X$-ray emission and DM direct detection searches. 

Concretely, along both these 3-to-2 and light mediator options  we followed a simple procedure.
We first check whether the ordinary $T=T'$ freeze-out scenario, either from the connector interaction or from the HS interaction, 
 could satisfy all constraints. As far as we know, in most models this does not work.  Second we investigate if the three regimes where both sectors do not thermalize with each other -- dark freeze-out with $T'<T$, reannihilation and freeze-in -- could solve these problems. For any model where DM belongs to a sector communicating with the SM through a portal, one could in principle  apply a similar procedure.

To perform the analysis on the basis of a concrete model we have considered the
HVDM model. 
This model involves only four free parameters, namely the vector DM mass $m_A$, the mediator mass $m_\eta$, the HS gauge interaction $\alpha_X$ and the Higgs portal interaction $\lambda_m$ (or equivalently the scalar mixing angle $\sin \beta$). 
As expected, within this model the solution of the small scale problems is incompatible with the ordinary freeze-out scenario.
Nevertheless two alternative production scenarios do work easily.
These scenarios are: $(a)$ dark freeze-out via 3-to-2 annihilation of a  HS colder than the SM, and $(b)$ freeze-in via the exchange of a light mediator.  

First let us summarize what happens along the various regimes for the case of no light mediator, i.e.~{\it considering a 3-to-2 annihilation}. 
In this case, the small-scale structure problems and the Higgs invisible decay constraints require sub-GeV DM with sizable self-interactions and a relatively small connector.
The outcome is

\begin{itemize}
\item \underline{Portal interaction freeze-out}: Since a sub-GeV DM particle can annihilate via the Higgs portal into a couple of SM particles  only at a very suppressed rate (due to the mixing angle and the Yukawa coupling), this scenario
is not viable. 

\item \underline{Dark freeze-out with $T=T'$}: A scenario with DM annihilations via 3-to-2 interactions, assuming that the DM temperature $T'$ and the SM temperature $T$ are equal, is not feasible because: $(i)$ one can not get simultaneously the observed relic density and significant self-interactions with perturbative $\alpha_X$ couplings (see Fig.~\ref{ma-alpha-som});
$(ii)$ at the moment of the DM freeze-out, the Higgs portal is too suppressed to maintain the kinetic equilibrium between the two sectors, giving rise to tension with the formation of cosmological structures.

\item \underline{Dark freeze-out with $T'<T$}: This scenario solves the previous two problems.
On the one hand, $T'<T$ implies a less populated HS and  thus smaller annihilation cross sections are needed to produce the observed relic density.
 Accordingly, perturbative values of $\alpha_X$ are compatible with both the DM abundance  constraint and the self-interaction hypothesis (e.g. Fig.~\ref{ma-alpha-som}).
On the other hand, if $T'<T$ at the moment of the freeze-out, structure formation constraints are naturally fulfilled, simply because DM is colder than the SM particles.
On top of that, this regime can also satisfy all other experimental constraints previously mentioned.
Fig.~\ref{fig:caseB} shows the way the relic density is obtained in this case as a function of $T$ and $T'$ for an example of successful set of parameters.
In particular, when the 3-to-2 annihilations become efficient, the HS heats up with respect to the SM sector. 
The parameter space allowed in this scenario is depicted in Fig.~\ref{ma-alpha-som2}, which corresponds to DM masses between a few keV and $\sim100$\,MeV.

\item \underline{Reannihilation}: This scenario requires that at the time when the HS interactions get Boltzmann suppressed (i.e. $T'\lesssim m_A$), the connector still creates HS particles. Due to the low DM mass range required by DM self-interactions, in the HVDM model reannihilation never takes place. This is because the (mixing angle and Yukawa suppressed) Higgs portal ceases to be active as soon as $T$ goes below $\sim 10$\,GeV. One could think about considering tiny values for $T'/T$ to overcome this difficulty but this would be incompatible with  the observed DM abundance.
It must be stressed however that for other kind of portals not suppressed at low temperatures (for instance a $Z'$ portal), the reannihilation could work  similarly to the way the dark freeze-out with $T'<T$ does.

\item \underline{Freeze-in}: This regime requires a HS that never thermalizes within itself. This implies a value of $\alpha_X$ which leads to self-interactions too feeble to  solve small-scale problems, unless one considers very light DM ($3$~keV $\lesssim m_A\lesssim10$~keV) at the verge of being too hot to be allowed.

\end{itemize}

The discussion of the regimes for the case with a light mediator differs in many aspects. In this case annihilations  proceed via 2-to-2 processes, and  the self-interaction hypothesis requires a mediator mass below $\sim 100$\,MeV with the DM particle several orders of magnitude heavier, which leads to the following conclusions:

\begin{itemize}
\item \underline{Portal interaction freeze-out}: This scenario does not work. 
On the one hand, for $\ma\lesssim 1$\,GeV the DM annihilates mainly into the HS and not into SM pairs as required in this scenario.
On the other hand, direct detection experiments exclude DM heavier that $\sim 1$\,GeV because of the strong enhancement of the DM-nucleon cross section (due to the lightness of the mediator).

\item \underline{Dark freeze-out with $T=T'$}: In this case DM annihilates through $AA\to\eta\eta$ or $A\eta$.
The thermalization condition $T=T'$ implies that the connector interaction should be sizable. However, direct detection bounds and cosmological observations require a much smaller connector,  basically excluding this regime.

\item \underline{Dark freeze-out with $T'<T$}: Unlike the case without light mediators,  most of the parameter space in this regime turns out to be excluded. Generally speaking this stems from two facts. First, this scenario predicts a relatively large annihilation cross section from the requirement that the HS thermalizes. Second it predicts  yields such that $Y_\eta\gg Y_A$. All this  leads to strong tensions with BBN, CMB and $X$-ray constraints. 
Details can be found in Appendix~\ref{sec:AppendixC}. 

\item \underline{Reannihilation}: This regime is analogous to the previous one, in the sense that it also requires a relatively large annihilation cross section and gives $Y_\eta\gg Y_A$.
As a consequence, as the previous regime it is excluded except in corners of the parameter space.

\item \underline{Freeze-in}: Finally, and unlike all the other light mediator regimes above, the freeze-in regime can easily work. This is due to the following  three facts. First, since this scenario is based on the assumption that the HS never reaches thermal equilibrium, it requires smaller values of $\alpha_X$ and $\sin\beta$, which allow to fulfill the direct detection constraints. Second, 
this regime decouples the $\alpha_X$-driven self-interaction  hypothesis from production of the DM, which is $\alpha_X$-independent.
Finally, unlike the dark freeze-out regimes, the freeze-in scenario naturally predicts $\Omega_\eta/\Omega_A\ll 1$, which permits to easily satisfy the $X$-ray and CMB constraints.
Fig.~\ref{fig:caseD} shows an example of successful freeze-in evolution of the DM relic density, together with the mass range needed in this case.

\end{itemize}

To sum up very shortly, the production of self-interacting DM in the early Universe can be
achieved by considering a DM sector which is not in thermal equilibrium with the SM bath.
We considered this possibility in the context of the particularly simple HVDM model. We are not aware of a simpler model that can simultaneously address all the issues discussed here. Moreover, many
of the conclusions we came to in this particular model would certainly remain valid for other
models where DM interacts with the SM via a portal.

\section*{Acknowledgments}
The authors would like to thank L. Lopez-Honorez, M. Tytgat and P. Tziveloglou for very useful discussions. N.B. is supported by the São Paulo Research Foundation (FAPESP) under grants 2011/11973-4 and 2013/01792-8. The work of C.G.C., T.H. and B.Z. is supported by the FNRS, by the IISN and by the 
Belgian Federal Science Policy through the Interuniversity Attraction Pole P7/37. 
B.Z. is also supported by the ``Investissements d'avenir, Labex ENIGMASS''.
 
\appendix
 
\section{Invisible Decay of the SM scalar}
\label{sec:Higgsdec}

The HS increases the number of Higgs boson decay channels.
The new decay modes are into $\eta$ scalars and DM particles.
The latter contribute to the invisible decay of the Higgs boson, whereas the former  does it partially if subsequently the $\eta$ scalars, in addition to decay to (visible) SM particles, also decay into DM. The corresponding decay rates are
\begin{eqnarray}
\Gamma(h\to AA)&=&\frac{3g_X^2}{128\pi}\sqrt{\mh^2-4\ma^2}\frac{12 \ma^4 - 4 \ma^2 \mh^2 + \mh^4}{\ma^2\mh^2}\sin^2\beta\,,\\
\Gamma(h\to\eta\eta)&=&\frac{1}{128\pi}\sqrt{\mh^2-4\meta^2}\left[\frac{2\meta^2+\mh^2}{\ma\mh v}\right]^2\left[g_X v\cos\beta+2\ma\sin\beta\right]^2\sin^2\beta\,.
\end{eqnarray}
Moreover, all the decay widths of the Higgs boson into SM particles at tree-level decrease by a factor $\cos^2\beta$ compared to the SM ones.
For the $\eta$ scalar, the decay width into SM particles at tree-level is $\Gamma(\eta\to\text{SM SM})=\Gamma_\text{SM}(h\to\text{SM SM})|_{\mh\to\meta}\cdot\sin^2\beta$.  As for the $\eta$ decay width into DM particle, if kinematically allowed, it is given by
\begin{equation}
\Gamma(\eta\to AA)=\frac{3g_X^2}{128\pi}\sqrt{\meta^2-4\ma^2}\frac{12 \ma^4 - 4 \ma^2 \meta^2 + \meta^4}{\ma^2\meta^2}\cos^2\beta\,.
\end{equation}
In this work, we adopt the bound on the invisible Higgs branching ratio and on $\cos^2\beta$ obtained in the analysis of Ref.~\cite{Bechtle:2014ewa}, which accounts for all the
relevant experimental information on the Higgs boson properties, considering simultaneously a universal modification of the Higgs boson couplings to SM particles and the possibility of an invisible Higgs decay. By doing this, it is found that
\begin{equation}
\cos^2 \beta \left[ 1 - \text{Br}_\text{inv} \right] \gtrsim 0.81 \;\;\; 
\hspace{15pt}\text{with}\hspace{15pt}
\text{Br}_\text{inv} = \text{Br}(h\to A\,A)+\text{Br}(h\to\eta\,\eta)\,\text{Br}(\eta\to A\,A)^2\,
\end{equation}
for the case of no light mediator, where we are concerned with $m_A, m_ \eta \ll m_h/2$. This in turn gives a very strong bound on the mixing angle
\begin{equation}
\sin\beta \lesssim  10^{-4}\sqrt{{1\over \alpha_X }\cdot  {m_A \over \text{GeV}}}. \label{invi:Higgs}
\end{equation}
The situation of the light mediator case is different. First, due to its lightness, the mediator $\eta$ only decays to SM particles. Consequently, the decay of Higgs to $\eta$ is in practice \textit{visible} unless $\eta$ decays dominantly to neutrinos or the lifetime of $\eta$ is so long that it decays outside the detector. Moreover, for $m_A \ge m_h/2$, the other decay channel of Higgs into HS particles, $h\to A\,A$, is kinetically forbidden.  Therefore, if $\eta$ decays dominantly to charged fermions or photons  within the detector and $m_A \ge m_h/2$, there would be no sizable invisible decay of the Higgs boson, and thus no relevant bound on $\sin\beta$. Otherwise, a bound similar to Eq.~\eqref{invi:Higgs} holds.


\section{Boltzmann Equations for the 3-to-2 Annihilation and the Instantaneous Freeze-out Approximation}
\label{sec:BEandFO}

In this appendix we assume that the dark and visible sectors have temperatures $T$ and $T'$ which are not necessarily equal. With this in mind, 
the evolution of the number densities $\n_i$ of the three DM particles $A_i$ are described by a system of three Boltzmann equations
{\small
\begin{eqnarray}
&&\frac{d\n_1}{dt}+3H\,\n_1=+\RR{1}{2}{3}{1}{1}\nonumber\\
&&-\RR{1}{2}{3}{2}{2}-\RR{1}{2}{3}{3}{3}\nonumber\\
&&-\RR{1}{1}{3}{1}{2}+\RR{2}{2}{3}{1}{2}\nonumber\\
&&+\RR{3}{3}{3}{1}{2}-\RR{1}{1}{2}{1}{3}\nonumber\\
&&+\RR{2}{2}{2}{1}{3}+\RR{2}{3}{3}{1}{3}\nonumber\\
&&-3\RR{1}{1}{1}{2}{3}-\RR{1}{2}{2}{2}{3}\nonumber\\
&&-\RR{1}{3}{3}{2}{3}\,,
\end{eqnarray}}
with two equivalent expressions for $\n_2$ and $\n_3$.
Here, $\neqq_i$ corresponds to the equilibrium densities and $\langle ijk\to lm\rangle\equiv\langle\sigma v^2\rangle_{ijk\to lm}$. 
 Due to the custodial symmetry the relation $\n\equiv \n_1=\n_2=\n_2$ holds and we find that 
$\NN{1}{2}{3}{1}{1}=\NN{1}{2}{3}{2}{2}=\NN{1}{2}{3}{3}{3}$, $\NN{1}{1}{1}{2}{3}=\NN{2}{2}{2}{1}{3}=\NN{3}{3}{3}{1}{2}$ and $\NN{1}{1}{3}{1}{2}=\NN{1}{1}{2}{1}{3}=\NN{2}{2}{3}{1}{2}=\NN{1}{2}{2}{2}{3}=\NN{1}{3}{3}{2}{3}=\NN{2}{3}{3}{1}{3}$. Thus the Boltzmann equations reduce to
\begin{equation}
\frac{d\n}{dt}+3H\,\n=\Big(-\NN{1}{2}{3}{1}{1}-\NN{1}{1}{1}{2}{3}-2\NN{1}{2}{2}{2}{3}\Big)\left(\n^3-\n^2\,\neqq\right)\,.
\end{equation}

As a result, total DM number density $N'=\n_1+\n_2+\n_3=3\,\n$ is given by
\begin{equation}
\frac{dN'}{dt}+3H\,N'=-\langle\sigma v^2\rangle_{3\to 2}\left(N'^3-N'^2\,{N'}^\text{eq}\right)\,,
\label{BE}
\end{equation}
where
\begin{equation}
\langle\sigma v^2\rangle_{3\to 2}\equiv\frac{\NN{1}{2}{3}{1}{1}+\NN{1}{1}{1}{2}{3}+2\,\NN{1}{2}{2}{2}{3}}{9}\,.
\label{sv32}
\end{equation}
 Following Ref.~\cite{Bernal:2015bla} to calculate the cross section in the non-relativistic limit, we find
\begin{equation}
\langle\sigma {v^2}\rangle_{3\to 2}=\frac{25\sqrt{5}\pi^2\,\alpha_X^3}{186624\ma^5}\,\frac{503295 \ma^8 + 592316 \ma^6 \meta^2 + 36188 \ma^4 \meta^4 - 156224 \ma^2 \meta^6 + 26209 \meta^8}{(\meta^2-4 \ma^2)^2 (\ma^2 + \meta^2)^2}\,.
\end{equation}

The Boltzmann Eq.~\eqref{BE} gives the DM number density as a function of time. Since we assume that the SM particles dominate the expansion of the Universe, we can trade the time variable $t$ for the temperature $T$ of the SM. By performing this change of variables we obtain
\begin{equation}
x\,H(x)\frac{dY(x)}{dx}=-s(x)^2\,Y(x)^2\,\langle\sigma v^2\rangle_{3\to 2} \, \big(Y(x)-Y^\text{eq}(x)\big)\,,
\label{BEY}
\end{equation} 
where $x\equiv \ma/T$ and $Y(x)\equiv N'/s(x)$. Notice that we define $Y(x)$ as a function of the SM quantities $x$ and $s(x)$, because it can be related to the today DM relic abundance by means of
\begin{equation}
\Omega_A = \frac{\ma\,N'}{\rho_c}=\frac{\ma\,s(\infty)\,Y(\infty)}{\rho_c}=\left(2.742 \cdot 10^8 \,\text{GeV}^{-1} \, h^{-2}\right) \, \ma \, Y(\infty)\,. 
\label{RelicDensity}
\end{equation}
 In this paper we use the following approximate solution of Eq.~\eqref{BEY}. 
After the freeze-out the DM density deviates from its equilibrium value. In fact,  as it becomes much greater, $ Y \gg Y^\text{eq}$, one can drop the $Y^{eq}$ term in Eq.~(\ref{BEY}), so that
\begin{equation}
x\,H(x)\frac{dY(x)}{dx}=-s(x)^2\,Y(x)^3\,\langle\sigma v^2\rangle_{3\to 2}\,.
\end{equation} 
This equation admits the following analytical solution
\begin{equation}
\frac{1}{Y(\infty)^2} = \frac{1}{Y(\xf)^2} + 2\int^\infty_{\xf} \frac{s^2\,\langle \sigma v^2\rangle_{3 \to 2} }{x\, H} dx\,.
\end{equation}
Here $\xf$ corresponds to the freeze-out temperature,  to be determined below. In general, one expects $ Y(\infty) \ll Y(\xf)$ and therefore 
\begin{equation}
Y(\infty) = \left[ 2 \int^\infty_{\xf} \frac{s^2\,\langle \sigma v^2\rangle_{3 \to 2} }{x\, H} dx\right]^{-1/2}\,.
\end{equation}
Now, since the DM particles are non-relativistic at freeze-out, the cross section $\langle \sigma v^2 \rangle_{3\to 2}$ is independent of the temperature.  It is therefore possible to conclude that in order to account for the observed DM relic density, Eq.~\eqref{RelicDensity}, one needs
\begin{equation}
\langle \sigma v^2 \rangle_{3\to2} \approx \left( 8.65\,\text{GeV}^{-5}\right) \,\xf^4 \, g_{*\text{FO}}^{-1.5} \left(\frac{1\, \text{GeV}}{\ma} \right)^2\,,
\label{FOconditionbis}
\end{equation}  
where $g_{*\text{FO}}$ corresponds to the number of relativistic degrees of freedom at the moment of the freeze-out.
 Note that it is the SM temperature at the freeze-out which enters in this formula and not the DM one. 

In order to estimate $\xf$, it is necessary to establish when the annihilation rate per particle  $(\neqq)^2 \langle \sigma v \rangle_{ 3\to 2}$ drops below the expansion rate of the Universe. Using Eq.~\eqref{FOconditionbis} it is found that this happens when freeze-out temperatures satisfy 
\begin{equation}
\xpf = 22.3 - \ln \left[\left(\frac{\xpf}{\xf}\right)^3\left(\frac{100 \,\text{MeV}}{\ma}\right)\left(\frac{g_{*\text{FO}}}{10}\right)\left(\frac{\xpf}{22.3}\right)^{-1.5}\right]\,.
\label{xpf}
\end{equation}
Injecting the value of $x'_\text{FO}$ one needs for a given value of the entropy ratio $\xi$ according to Eq.~\eqref{3to2:decouple}, this equation gives the value of $x_\text{FO}$.
\section{Constraints on Regimes 2 and 3A with a light mediator}
\label{sec:AppendixC}

In the scenario with a light mediator, two essential features of regimes 2 and 3A are that both lead to $Y_\eta \gg Y_A$ and relatively large annihilation cross sections. On the one hand, the former feature arises because the abundance of $A$ undergoes a Boltzmann suppression in contrast to the one of the lighter $\eta$ scalar. On the other hand, the latter feature is necessary in order for the HS sector to thermalize with itself. Based on these two observations, we consider the constraints on regimes 2 and 3A by following Fig.~\ref{fig:Regime3A}. 

First we consider the cases in which the $\eta$ scalar annihilates into an electron-positron pair, that is, $m_\eta>2\,m_e~\sim 1$\,MeV.  It is convenient to discuss them according to the value of the lifetime of $\eta$.

\begin{figure}[t]
\vspace{-.5cm}
\begin{center}
\includegraphics[width=0.8\textwidth]{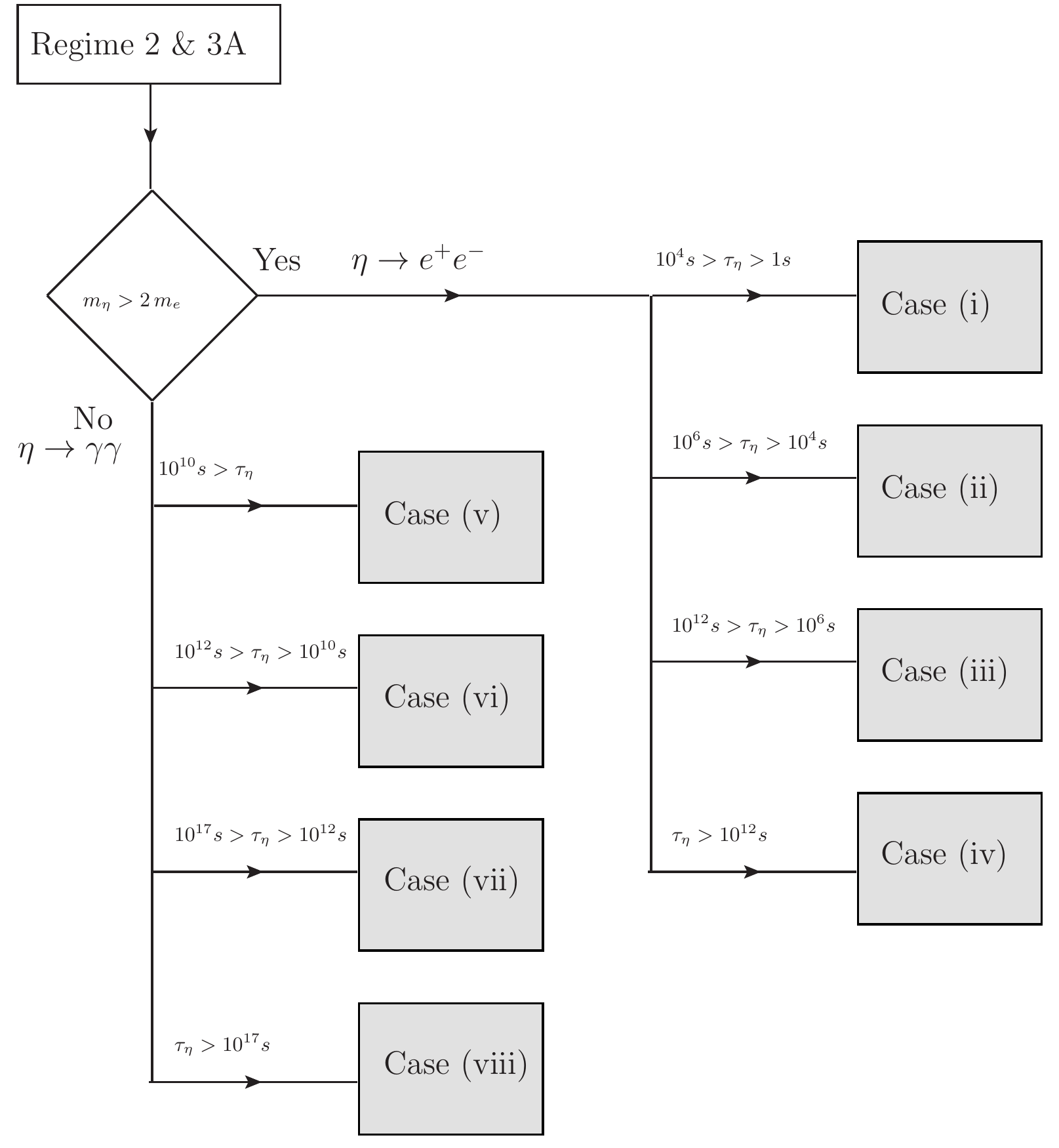}
\end{center}
\vspace{-.5cm}
\caption{ Flow chart illustrating the phenomenology which applies to the regimes 2 and 3A, depending on the values of $m_\eta$ and of the lifetime of the $\eta$ particle.}
\label{fig:Regime3A}
\end{figure}

\begin{itemize}
\item[$(i)$] $10^4$~s $>\tau_\eta$:  Here, BBN bounds do not apply because the $\eta$ particle has a short lifetime. Nonetheless, the requirement of non-thermalization between both sectors along with DM direct detection searches excludes this case. For illustration, we show this in the left panel of Fig.~\ref{fig:Regime3Ai}, which corresponds to $m_\eta=10$\,MeV. Although we take $\langle\sigma v\rangle_{AA\to\eta\eta} = 3\cdot 10^{-28}$ cm$^{3}/$s for concreteness, our conclusions are general.

Note that for a DM mass of $\mathcal{O}$(10)\,GeV, by increasing $m_\eta$ beyond 10\,MeV, the direct detection constraints get relaxed but this possibility is excluded because such a mass spectrum is not compatible with the required self-interactions, as shown in the right panel of Fig.~\ref{fig:Regime3Ai}.

\item[$(ii)$] $10^6$~s $>\tau_\eta>10^{4}$~s: The mixing angle $\beta$ corresponding to this lifetime is still quite large, which leads to $\eta$ abundance larger than 1. The late decay of the $\eta$  scalar can spoil the predictions of BBN~\cite{Jedamzik:2009uy}. As a result, this possibility  is basically excluded. 

\item[$(iii)$] $10^{12}$~s $>\tau_\eta>10^{6}$~s: In contrast to the previous case, the mixing angle required  here does not necessarily imply a huge $\eta$ abundance. Still in order to satisfy the BBN bound, $\Omega_\eta\lesssim 10^{-5}$ is needed. Since in this scenario $Y_\eta \gg Y_A$, this constraint and $\Omega_A h^2 \sim 0.12$ lead  to a $m_A$ larger than $10^4\,m_\eta$ by orders of magnitude. However for such large DM masses, in order to have significant self-interactions to address small scale problems, one needs a very small mediator mass, as can be seen in the right panel of Fig.~\ref{fig:Regime3Ai}. Therefore, in practice it is hardly allowed to have $m_\eta > 1$\,MeV and $\Omega_\eta\lesssim 10^{-5}$ simultaneously. As a result, this case is strongly disfavored.

\begin{figure}[h]
\begin{center}
\includegraphics[width=0.49\textwidth]{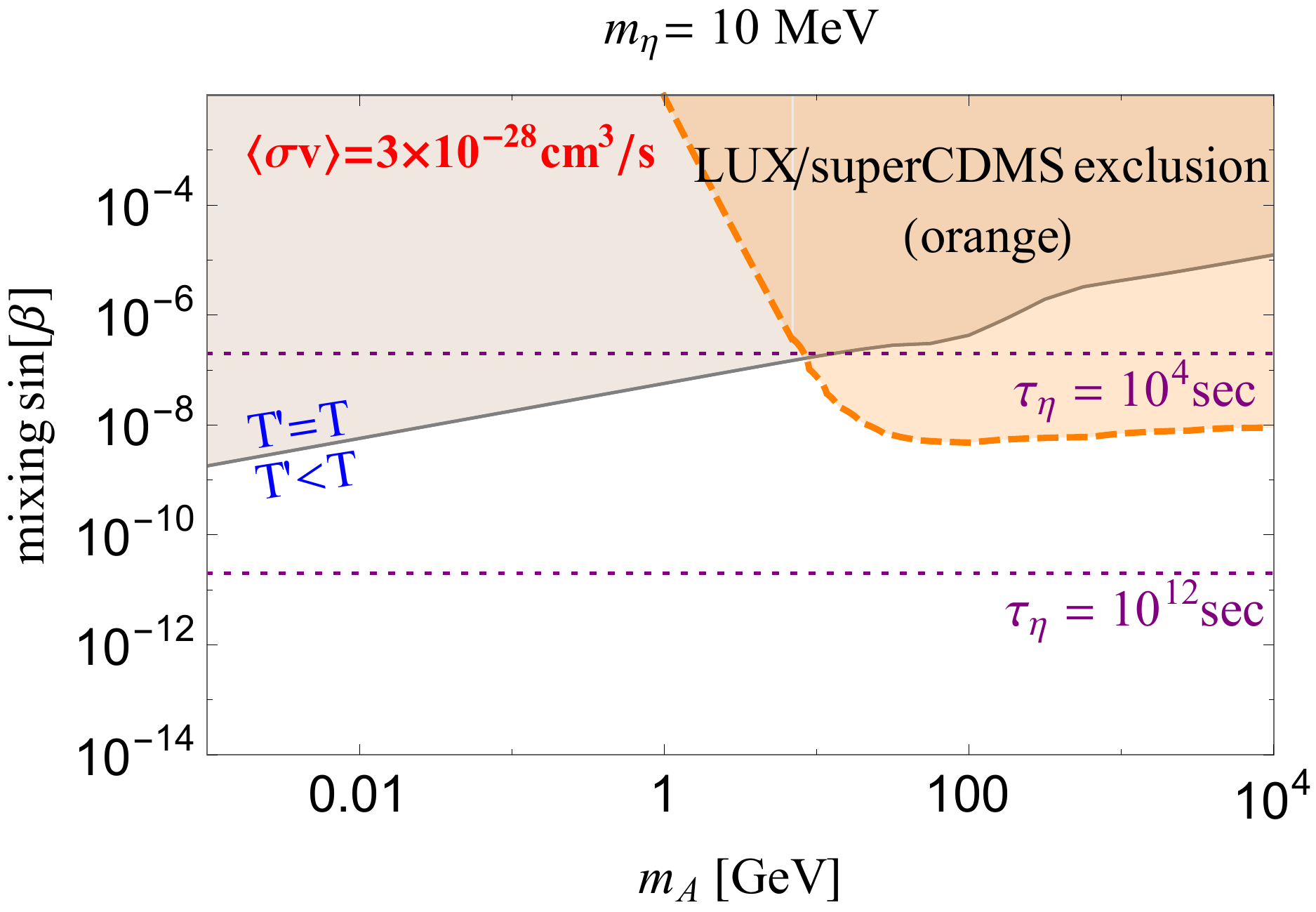}
\includegraphics[width=0.49\textwidth]{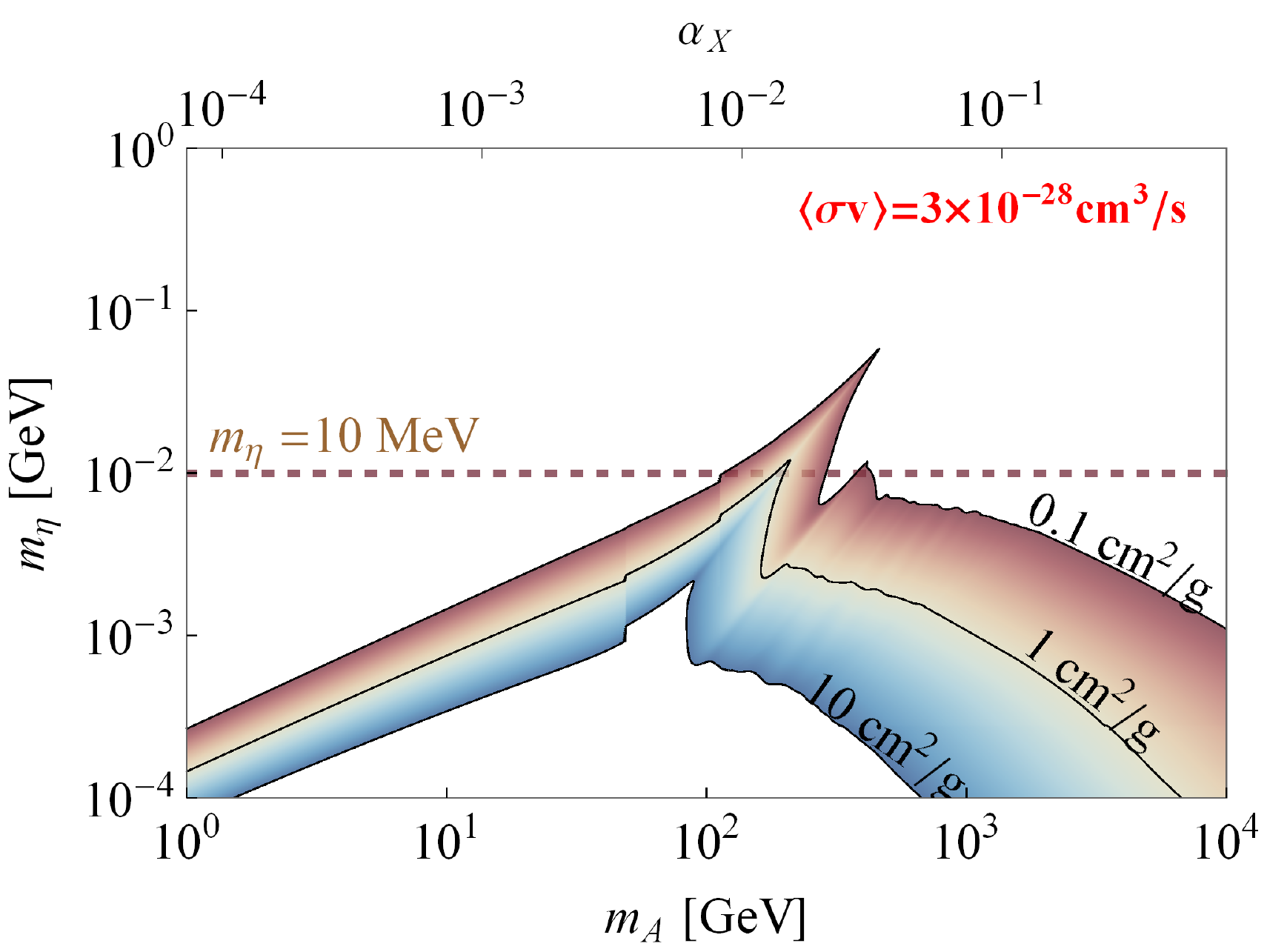}
\end{center}
\vspace{-0.5cm}
\caption{Left panel: constraints on regimes 2 and 3A for $m_\eta>2\,m_e$. The orange region is excluded by direct detection searches.  The light-gray shaded region gives $T=T'$, i.e.~corresponds to regime 3B. Here we take $m_\eta=10$\,MeV and $\langle\sigma v\rangle_{AA\to\eta\eta} = 3\cdot 10^{-28}$ cm$^{3}/s$. Right panel: Dark matter self-interactions for the regimes 2 and 3A, yielding $\sigma_T/\ma$ in the range of $0.1-10$ cm$^2$/g (top-down) at $v\sim 10$~km/s. The upper horizontal axis gives the value of $\alpha_X$ which leads to $\langle\sigma v\rangle_{AA\to\eta\eta} = 3\cdot 10^{-28}$ cm$^{3}/s$.}
\label{fig:Regime3Ai}
\end{figure}

\item[$(iv)$] $\tau_\eta>10^{12}$~s:  If we assume, as we do here, that the HS is mostly created from the visible sector, $\sin \beta$ can not be arbitrarily small. In this case, it can be shown that $\tau_\eta<10^{17}\,s$ must hold (if the $\eta$ decays into electrons as assumed here). As a result the decay of $\eta$ can modify the CMB spectrum, which leads to the following constraint: $\Omega_\eta\lesssim 10^{-8}$. Using similar arguments as in case $(iii)$, this requires even much larger DM masses and it is therefore basically excluded. 
\end{itemize}

Next we consider the cases in which the $\eta$ scalar decays into photons, that is, $m_\eta\lesssim$ 1\,MeV. 
\begin{itemize}
\item[$(v)$] $10^{10}$~s $>\tau_\eta$: Similarly to case $(i)$, the requirement  of non-thermalization of the HS with the visible sector, together with DM direct detection constraints, exclude this possibility. We illustrate this scenario in Fig.~\ref{fig:Regime3Av}, for $m_\eta=0.003$\,MeV and 0.5\,MeV, respectively. We assume $\langle\sigma v\rangle_{AA\to\eta\eta} = 3\cdot 10^{-28}$~cm$^{3}/$s in both cases.

\begin{figure}[h]
\begin{center}
\includegraphics[width=0.49\textwidth]{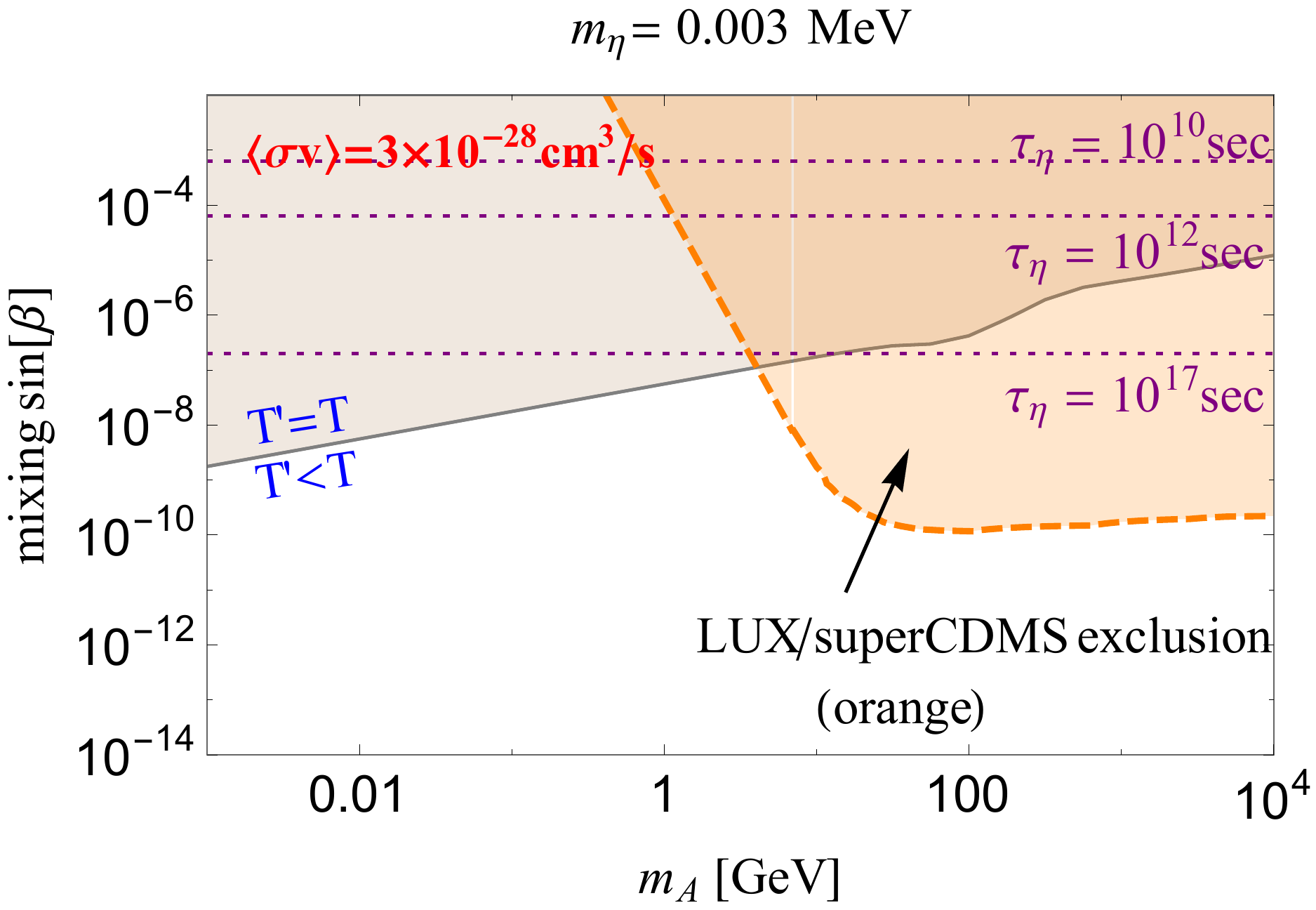}
\includegraphics[width=0.49\textwidth]{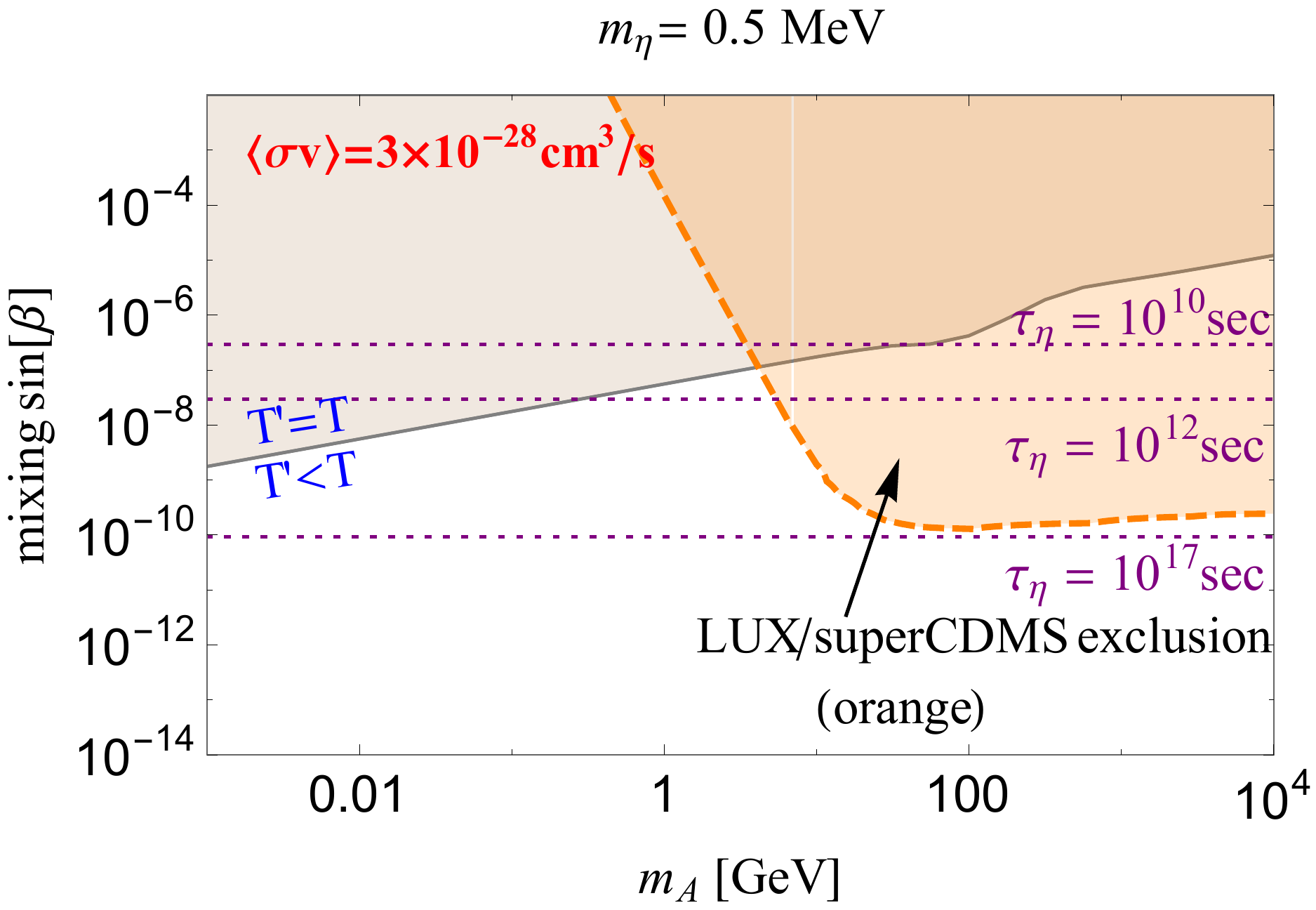}
\end{center}
\vspace{-0.5cm}
\caption{Constraints on regimes 2 and 3A for the $m_\eta < 2\,m_e$ case,  for both a very light ($m_\eta=0.003$\,MeV) or a relatively heavy ($m_\eta=0.5$\,MeV) $\eta$. The exclusion regions are defined in the same way as in Fig.~\ref{fig:Regime3Ai}. The value of $\langle\sigma v\rangle_{AA\to\eta\eta}$ has been taken to be $3\cdot 10^{-28}$ cm$^{3}/$s for both panels.
}
\label{fig:Regime3Av}
\end{figure}

\item[$(vi)$] $10^{12}$~s $>\tau_\eta>10^{10}$~s: This part of the parameter space is not as severely constrained as the previous cases. Nevertheless, we find only a small parameter region that satisfies both the relic abundance constraint and the self-interaction expectation, namely 1\,GeV $<m_A< 10$\,GeV, 0.1\,MeV $<m_\eta<2\,m_e$ and $10^{-7}< \sin\beta <10^{-8}$. One example is given in right panel of Fig.~\ref{fig:Regime3Av} for $m_\eta = 0.5$\,MeV. While this scenario might be further constrained by searches of axion-like particles (ALPs),  we will not enter into the details of this analysis here.
\item[$(vii)$] $10^{17}$~s $>\tau_\eta>10^{12}$~s: Considering  the direct detection bounds and requiring that the HS does not thermalize with the visible sector, from Fig.~\ref{fig:Regime3Av} one can see that this scenario is only consistent with relatively heavy $\eta$.
Similarly to case $(iv)$, CMB constraints apply, leading to DM masses much larger than 100\,GeV. Such parameters are strongly constrained by direct detection bounds and lead to tiny mixing angles, of the order of $10^{-10}$ (see right panel of Fig.~\ref{fig:Regime3Av} for an example).  These constraints typically require
parameters which lead to a freeze-in scenario, rather than a scenario where the HS thermalizes with itself. This possibility is therefore excluded.
\item[$(viii)$] $\tau_\eta>10^{17}$~s:  In this case $\eta$ is long-lived and the $X$-rays constraint of Eq.~\eqref{caseD:Xray} applies.  If again we assume that the HS is mostly created from the visible sector, so that $\sin \beta$ can not be arbitrarily small, the $X$-ray constraint implies that $m_\eta \lesssim 1$~keV. For DM masses above keV, such light mediators would strongly enhance DM self-interactions in contradiction with astrophysical observations~\cite{Clowe:2003tk,Markevitch:2003at,Randall:2007ph}. This case is therefore excluded. 
\end{itemize}
\bibliographystyle{JHEP}
\bibliography{biblio}

\end{document}